\documentclass[aps, prd, showpacs, nofootinbib, preprintnumbers,floatfix,longbibliography, reprint, superscriptaddress]{revtex4-1}
\usepackage[utf8]{inputenc}
\usepackage[colorlinks=true, linkcolor=blue, citecolor=teal]{hyperref}
\usepackage{bm}
\usepackage{latexsym,amssymb,amsmath,amsfonts}
\usepackage{graphicx}
\usepackage{physics}
\usepackage{bm}
\usepackage{footnotebackref}
\usepackage{multirow}
\usepackage[usenames]{xcolor}
\usepackage{soul}
\usepackage{makecell}
\usepackage{graphicx}
\usepackage{amssymb}
\usepackage{amsmath}
\usepackage{rotating}
\usepackage{color}
\usepackage{fancyhdr}
\usepackage{ifthen}
\usepackage{slashed}
\usepackage{xspace}
\usepackage{tabularx}
\usepackage{float}
\usepackage{subcaption}

\usepackage{tikz}
\usetikzlibrary{shapes.geometric, arrows}
\tikzstyle{startstop} = [rectangle, rounded corners, minimum width=3cm, minimum height=1cm,text centered, draw=black, fill=red!30]
\tikzstyle{io} = [trapezium, trapezium left angle=70, trapezium right angle=110, minimum width=3cm, minimum height=1cm, text centered, draw=black, fill=blue!30]
\tikzstyle{process} = [rectangle, minimum width=3cm, minimum height=1cm, text centered, draw=black, fill=orange!30]
\tikzstyle{decision} = [diamond, minimum width=1cm, minimum height=1cm, text centered, draw=black, fill=green!30]
\tikzstyle{arrow} = [thick,->,>=stealth]

\newcommand{\hatom}{\hat{\Omega}}
\newcommand{\al}{\alpha}
\newcommand{\intd}{{\rm d}}
\newcommand{\hpx}{{\tt HEALPix}}
\newcommand{\pyStoch}{{\tt PyStoch}}

\begin{document}

\title{Efficient search for detection candidates using a peak finder strategy for all-sky-all-frequency gravitational wave radiometer}

\author{Arindam Sharma}
\email{asharm17@go.olemiss.edu}
\affiliation{Department of Physics and Astronomy, University of Mississippi, University, Oxford, Mississippi 38677, USA}

\author{Deepali Agarwal}
\email{deepali.agarwal@utrgv.edu}
\affiliation{Department of Physics and Astronomy, University of Texas Rio Grande Valley,\\
One West University Boulevard, Brownsville, Texas 78520, USA}
\affiliation{Centre for Cosmology, Particle Physics and Phenomenology (CP3),\\ Universite catholique de Louvain, Louvain-la-Neuve, B-1348, Belgium}
\affiliation{Inter-University Centre for Astronomy and Astrophysics (IUCAA), Pune 411007, India}

\author{Sanjit Mitra}
\email{sanjit@iucaa.in}
\affiliation{Inter-University Centre for Astronomy and Astrophysics (IUCAA), Pune 411007, India}

\begin{abstract}

The first all-sky-all-frequency (ASAF) radiometer search was conducted using data from the first three observing runs of the Advanced LIGO and Advanced Virgo detectors. The significance of this search lies in its fast and unmodeled approach, leveraging a cross-correlation technique to identify common signals across the detector network. As a result, this method serves as an excellent alternative to search for unknown or poorly modeled continuous wave sources and narrow band components of the gravitational wave (GW) background.
For continuous wave sources whose waveform can be modeled, this method can serve as the first stage of a hierarchical scheme by identifying subthreshold candidates to be followed up with more optimal but computationally expensive searches. The ASAF search, however, presently suffers from beam smearing, where multiple candidates may arise due to the same noise fluctuations, detector artifact, or a GW source. This can reduce the detection probability in follow-up analyses, especially with limited computing resources. To mitigate this issue and reduce the number of correlated and unnecessary candidates, we introduce a novel peak finder algorithm. This algorithm helps identify the most representative candidates while preserving detection sensitivity, thereby allowing follow-up of a much larger number of independent candidates. This reduction in correlated samples leads to a significant reduction in the number of trial factors and also the false dismissal rate (FDR), depending upon the frequency, strength of the injected signal, and the number of candidates, $N$, that can be followed up, with the reduction in FDR being most significant at low frequencies, and small $N$. For instance, at 30 Hz, following up $N=2$ peak finder candidates reduces FDR by a factor of $\sim3$. At high $N$, the peak finder FDR is higher than the full sky FDR. However, this is of no practical consequence as we can only follow up a small number of candidates per frequency bin.  

\end{abstract}

\maketitle

\section{Introduction}\label{sec:intro}
The gravitational wave (GW) transient catalog~\cite{LIGOScientific:2025slb} recently released by the LIGO-Virgo-KAGRA (LVK) Collaboration, featuring more than 200 sources, marks a stage where merging black holes and neutron stars are being routinely observed. However, aside from these transient events, persistent signals, such as continuous gravitational waves (CWs) and gravitational wave backgrounds (GWBs) are yet to be detected in the audio-frequency band [refer to the reviews~\cite{universe5110217,universe7120474,Piccinni:2022vsd,Riles:2022wwz,Wette:2023dom,Haskell:2023yrv} and~\cite{Allen:1996vm,Christensen:2018iqi,sym14020270,Caprini:2018mtu,Romano_rev} for sources and detection methods of CWs and GWB respectively, and references therein].

The CWs are everlasting quasiperiodic GWs which may be generated by spinning compact objects with deviations from axial symmetry in their distribution of mass and energy. The search sensitivity for CW searches has started surpassing the spin-down limits~\cite{Riles:2022wwz} for several known pulsars, and their astrophysically motivated parameter space is being constrained via GW observations, though such signals have not been detected so far~\cite{Abbott_2021,Abbott_2022}. 

GWBs are persistent signals expected to arise from the superposition of individually undetectable or unresolved astrophysical GW sources~\cite{sym14020270}, as well as from phenomena in the early Universe, such as inflation, first-order phase transitions, and cosmic strings~\cite{Allen:1996vm,Caprini:2018mtu,Christensen_2019}. The waveform is unpredictable for GWB due to a large number of contributing sources with unknown parameters~\cite{CornishRomano}, making a matched-filtering-based search infeasible. Searches for GWB rely on cross-correlating data from geographically separated detectors looking for a common excess power~\cite{AllenRomano,Romano_rev}. A GWB can be categorized according to spectral and angular distributions, such as narrow band or broad band, and isotropic or anisotropic (pointlike or extended sources). 

Searches for anisotropic GWB rely on creating ``maps" of the cross power distribution by utilizing the direction dependent time delay between detectors~\cite{Allen:1996gp,Ballmer_2006,Radiometer_Mitra_2008}. This method, known as the GW radiometer algorithm, is essentially aperture synthesis using the Earth's rotation. The LVK Collaboration has applied this method on data up to the first part of the fourth observing run (O4a) and set upper limits on directional flux and angular power spectrum of anisotropic GWBs~\cite{PhysRevD.76.082003,PhysRevLett.107.271102,PhysRevLett.118.121102,O1O2Radiometer,O3Aniso,LIGOScientific:2025bkz}. 

The radiometer algorithm has the capability of detecting CW sources too~\cite{PhysRevD.76.082003, PhysRevLett.107.271102, Messenger2015, PhysRevLett.118.121102, O1O2Radiometer, O3Aniso, MeyersThesis, Salvadore:2025lba}. 
As an unmodeled search, this method is robust to GWs from poorly understood CW sources, such as accreting neutron stars in x-ray binaries, and exotic sources, such as boson clouds around spinning black holes. A targeted narrow band radiometer search, where the radiometer is ``pointed'' to a specific direction and a narrow band GW signal is searched over the sensitive frequency range for terrestrial detectors (i.e., 20–1726 Hz), has been conducted in the past by the LVK Collaboration~\cite{PhysRevD.76.082003,PhysRevLett.107.271102,PhysRevLett.118.121102,O1O2Radiometer,O3Aniso,LIGOScientific:2025bkz}. This search set constraints on the GW strain from potential CW source directions such as Scorpius X-1, SN 1987A, and the Galactic Center.

The first all-sky-all-frequency (ASAF) radiometer search was conducted during the third LVK observing run (O3)~\cite{ASAF_O1-O3}. This was made possible by overcoming computational constraints through the use of data folding techniques~\cite{Ain_2015} and the \pyStoch~\cite{Ain_2018} pipeline, enabling the creation of sky maps on the \hpx{} grid\footnote{http://healpix.sourceforge.net}~\cite{Górski_2005,Zonca}. The \hpx{} grid parameters were set to {\tt nside}=16, corresponding to 3072 sky pixels and pixel size of $\sim$ 13.4 degrees. No evidence of a pointlike narrow band persistent source has been found with the latest O4a results, and instead, upper limits on the effective strain amplitude have been set~\cite{LIGOScientific:2025bkz}.

The capability of the ASAF search can be further augmented by identifying detection candidates and performing follow-up analyses using more sensitive search pipelines more optimal to search for CW sources. A total of 515 subthreshold candidates identified by the ASAF search were followed up using the Viterbi algorithm (which allows for spin wandering)~\cite{Knee2024}. In this hierarchical approach, the search sensitivity highly depends on the number of reliable follow-up candidates.

The candidates in the ASAF search are nothing but frequency-pixel pairs. One of the challenges in this search is the large sample size, which happens to be approximately $\sim 10^{8}$ for {\tt nside}=16 and about 50,000 frequency bins. With the available computing resources, only a few hundred top candidates can be followed up. To reduce the loss in signal-to-noise (SNR) ratio caused by mismatches between the actual direction of a point source and the center of the closest pixel, the sky maps are created at a sky resolution sufficiently finer than the minimum beam size at that direction. Owing to fine pixels, the highly anisotropic and extended nature of the beam, the detection statistic across the pixels may not be independent. The point spread function~\cite{Radiometer_Mitra_2008} or the beam varies with signal frequency, the baseline length, and sky direction. The minimum width of the primary lobe of the beam (essentially the \textit{diffraction limit}) for the baseline constituted by the LIGO Hanford (H) and Livingston (L) detectors has a width $\Delta \theta_{\rm Diff}$ (in radians)~\cite{PhysRevLett.118.121102,Floden22}
\begin{equation}\label{eq:diffraction}
    \Delta \theta_{\rm Diff} \approx \frac{50}{f\,({\rm Hz})}\,,
\end{equation}
where $f$ is the signal frequency. 

\begin{table}[t]
    \caption{Approximate beam width (diffraction limit) and number of correlated pixels (in the diffraction limited region) for different frequencies with the LIGO Hanford-Livingston baseline. The exact numbers depend on the sky direction. In this case, we have chosen (RA, Dec) = $(6.2\ {\rm h}, 45^\circ)$.
    }
    \begin{ruledtabular}
    \begin{tabular}{c|ccccc}

                        $f$ (Hz) & 30 & 100 & 200 & 400 & 700  \\
                            \hline
                 $\Delta \theta_{\rm Diff}$ (degrees) & 95.5 & 28.6  & 14.3   & 7.2   & 4.1    \\
    No. of correlated pixels & 503   & 49     & 12      & 3      & 1
    \end{tabular}
    \end{ruledtabular}
    \label{table:size_diff_lim_region}
\end{table}

We have listed the beam sizes in Table~\ref{table:size_diff_lim_region} for a range of sample frequencies. Since the pixel resolution has to be chosen to avoid significant SNR losses across a broad frequency band, the number of pixels correlated at the lower frequencies can be very high, as shown in Table~\ref{table:size_diff_lim_region}. The number of correlated pixels is calculated using the {\tt query\_disc} function from the {\tt healpy} package with an example noisy sky map (without source injection). The {\tt query\_disc} function returns the pixels whose centers are within the disc defined by a vector and radius, chosen to be (RA, Dec) = $(6.2\,{\rm h}, 45^\circ)$ and $\Delta \theta_{\rm Diff}/2$, respectively. Then, we count the number of pixels inside the disc, which are mentioned in Table~\ref{table:size_diff_lim_region}. 

Consequently, a large number of candidates (pixels) can be triggered due to the same cause (spurious detector disturbance or GW source). One common way to address this issue is to employ clustering in sky coordinate parameter space. The clustering algorithm may vary, but the core idea is to identify candidates arising from the same root cause, club them together, and treat them as a single entity during the follow-up stage. The clustering procedure enhances the detectability of a GW source by freeing up computational resources and allowing candidates with lower significance to be followed up. Various clustering algorithms have also been employed in the preprocessing stage of CW search pipelines~\cite{DirectedCW_GC2013,MAP_2015,MAP_2017,MAP_2021,MAP_2022}.

Here, we explore a peak-finder-based clustering statistic, where we identify the ``peaks" in the sky maps for follow-up studies. A peak is defined as a pixel whose SNR exceeds that of all its nearest-neighbor pixels. The SNR values associated with these peak pixels, referred to as \textit{peak SNR}, are employed as a detection statistic. We test our peak SNR statistic with simulated datasets and demonstrate the improvement in detection probability compared to the ``full-sky map" statistic, where all the pixels in the sky are utilized to identify candidates for follow-up studies.

The article is structured as follows: Sec.~\ref{sec:ASAF_methods} presents the formalism used in the ASAF search. In Secs.~\ref{sec:peak_finder} and~\ref{sec:sim_study}, we detail the peak finder algorithm and statistical characterization through Monte Carlo simulations, respectively. We conclude the article with a summary and a discussion of the future prospects of the methods proposed here in Sec.~\ref{sec:Conclusion}.

\section{All-Sky-All-Frequency Search} \label{sec:ASAF_methods}

The ASAF radiometer search aims to detect persistent GW sources localized in both the sky and frequency. In this section, we first review the formalism used to characterize and search for narrow band pointlike GWBs. We then discuss the similarities and differences between an anisotropic narrow band GWB and a CW source. Finally, we discuss the ``full-sky map'' detection statistic and its drawbacks at the end of this section.

\subsection{Characterization of narrow band anisotropic GWB}

The metric perturbations at a space-time point $(\vec{x},t)$, in transverse-traceless gauge, can be expanded in terms of plane waves incoming from the $\hatom$ direction and solid angle $\intd^2\hatom$, with frequency in range $f$ and $f+\intd f$, as~\cite{AllenRomano}
\begin{equation}
\begin{aligned}
   h_{ab} (t, \vec{x}) =\sum_{A=+,\times} &\int_{-\infty}^\infty \intd f \int_{S^2} \intd^2\hatom \\
    &\,e^A_{ab} (\hatom)\, \tilde{h}_A(f, \hatom)\,e^{i2\pi f (t - \hatom.\vec{x})/c}\,,
    \end{aligned}
\end{equation}
where $c$ is the speed of light. $\tilde{h}_A(f, \hatom)$ are Fourier domain strain coefficients and due to the real nature of signal $\tilde{h}_A(f, \hatom)=\tilde{h}^*_A(-f, \hatom)$. The index $A$ and tensor $e^A_{ab}$ represent the state of polarization and polarization tensor of GWs. 

The strain coefficients $\tilde{h}_A(f, \hatom)$ are, in principle, unknown for any kind of GW source. For instance, $\tilde{h}_A(f, \hatom)$ for a single monochromatic GW source will be proportional to $\delta (f-f_0)$, but deterministic. That is, one can design a grid of astrophysically motivated $f_0$ and search for a signal with frequency $f_0$ employing matched-filtering-based techniques, comparing the data stream to a set of templates in order to obtain high correlation. In practice, however, the parameter space to be explored is significantly larger due to uncertainties in quantities such as frequency, frequency derivatives and sky directions, etc. A ``blind'' fully coherent matched filtering-based search is computationally prohibitive and has been performed using suboptimal methods, e.g., via semicoherent and hierarchical schemes~\cite{universe7120474}.

On the other hand, if there is a superposition of a large (or infinite) number of GW sources with a statistical distribution of parameter values, the strain coefficients for the GWB are no longer predictable, we need to consider an ensemble of realizations for $\tilde{h}_A(f, \hatom)$. Assuming that the GW signal is statistically stationary and unpolarized, and $\tilde{h}_A(f, \hatom)$ is normally distributed, the first two moments of $\tilde{h}_A(f, \hatom)$ can be written as~\cite{AllenRomano},
\begin{equation}
\begin{aligned}
    \langle \tilde{h}_A(f, \hatom)\rangle_h &= 0\,, \\
    \langle \tilde{h}_A(f, \hatom)\,\tilde{h}^*_{A'}(f', \hatom')\rangle_h &= \\
    \frac{1}{4} \,\delta (f-f')&\,\delta^2(\hatom-\hatom') \,\delta_{AA'} \, \mathcal{P}(f,\hatom) \,,
\end{aligned}
\end{equation}
where $\mathcal{P}(f,\hatom)$ is the (polarization-averaged) one-sided power spectral density for GWB. $\mathcal{P}(f,\hatom)$ is the quantity of interest and is observable through GW experiments. The directional dependence of $\mathcal{P}(f,\hatom)$ could be well localized (into a pixel) or extended (over a large-scale structure). In such cases, $\mathcal{P}(f,\hatom)$  could be decomposed into a suitable basis as
\begin{equation}
    \mathcal{P}(f,\hatom) = \sum_\alpha \,\mathcal{P}_\alpha(f) \,e_\alpha(\hatom)\,,
\end{equation}
where $e_\alpha(\hatom)$ are basis vectors: 1 for isotropic source, and $e_\alpha(\hatom)=\delta^2 (\hatom-\hatom_\alpha)$ for a point source in direction $\hatom_\alpha$ and $e_\alpha(\hatom)=Y_{\ell m}(\hatom)$, where $Y_{\ell m}(\hatom)$ are spherical harmonic basis functions. The $\mathcal{P}_\alpha$ values are interpreted as the GWB power in the $\alpha$-th component. The primary goal of anisotropic GWB searches is to estimate $\mathcal{P}_\alpha(f)$, for which the usual strategy is to employ the GW radiometer algorithm~\cite{Ballmer_2006,Radiometer_Mitra_2008}. In this article, we will focus on mapping anisotropy of the GWB in a pixel basis. (Readers interested in the ASAF search in the spherical harmonic basis may refer to~\cite{ASAF_SpH}.) 

\subsection{Map making with GW radiometer}~\label{sec:Radiometer}

The GW radiometer algorithm uses cross correlation of data from geographically separated observatories. It is a semicoherent search, in which data from the full observing runs is divided into small time segments and the correlation is performed segmentwise before combining with optimal weighting factors. We briefly outline the process of constructing estimators for $\mathcal{P}_\alpha(f)$ using this algorithm. The time-series data from the observing runs are divided in $\tau=192$ s long, 50\% Hann-windowed overlapping time segments~\cite{ASAF_O1-O3,LIGOScientific:2025bkz}. The search begins by constructing the cross-spectral density (CSD), $C^I (t,f)$, as~\cite{Ain_2015,PhysRevD.104.022004},
\begin{equation}
   \bm{C}_f\equiv  C^I(t,f) \equiv \frac{2}{\tau\, \overline{w_1 w_2}}\,\tilde{s}_{I_1}^*(t,f) \,\tilde{s}_{I_2}(t,f)\,,
\end{equation}
where $ t $ is the time stamp of a given segment, and $ \overline{w_1 w_2}$ are window factors. $\tilde{s}_{I_i}(t,f)$ represents the short-term Fourier transform of data from detector $ I_i $ in the detector baseline $ I $. 

Assuming the detector noise is additive, Gaussian, and stationary within each segment, and uncorrelated with the noise of a geographically separated detector, it can be shown that~\cite{thrane2009,Ain_2015},
\begin{equation}
   \langle C^I (t,f) \rangle_{h,n} =  \sum_\al \mathcal{P}_\al(f)\, \gamma^I_\al (t,f) \,,
\end{equation}
where $\langle \,\cdot\, \rangle_{h,n}$ denotes the ensemble average over noise and GW source realizations. $\gamma^I_\al (t,f)$ is a detector and source geometry-dependent function, known as the overlap reduction function (ORF), given in pixel basis by~\cite{Radiometer_Mitra_2008},
\begin{equation}
\begin{aligned}
   \bm{\gamma}_f&\equiv \gamma^I_\al (t,f) \\
   &= \sum_A \frac{1}{2}\, F^A_{I_1}(\hatom_\alpha, t)\, F^A_{I_2}(\hatom_\alpha, t) \,e^{i2\pi f (\hatom_\alpha.\Delta\vec{x}_I(t))/c}\,,
    \end{aligned}
\end{equation}
where $\Delta\vec{x}_I(t)=\vec{x}_{I_1}(t)-\vec{x}_{I_2}(t)$  is the separation vector between the two detectors, and $F^A_{I_i}(\hatom, t)$ represents the detector response for detector $I_i$.

Given the persistent nature of the GW signal of interest, observations can be integrated over the entire observing run. This offers two advantages:
(i) The SNR improves as more data are accumulated and
(ii) sky coverage increases, as detector baselines have blind spots that shift across the sky due to Earth's rotation~\cite{Radiometer_Mitra_2008}.  

Assuming that the CSD for each time segment is a Gaussian random variable, the maximum likelihood (and unbiased) estimator for observable GWB anisotropy, $\mathcal{P}_\al(f)$, is obtained by maximizing the joint likelihood across positive and negative frequency, different times and baselines~\cite{ASAF_O1-O3}, 
\begin{equation}
    \bm{\hat{\mathcal{P}}}_f = \bm{\Gamma}_f^{-1}\, \cdot\, \bm{X}_f\,.
\end{equation}
Here, $\bm{X}_f$ is the narrow band \textit{dirty map} and $\bm{ \Gamma}_f$ is the narrow band Fisher information matrix defined as follows \citep{Ain_2018,ASAF_O1-O3},
\begin{equation}
\begin{aligned}
   \bm{X}_f&= \bm{\gamma}_f^\dagger \cdot \bm{N}_f^{-1} \cdot \bm{C}_f\,,\\
   \bm{\Gamma}_f&= \bm{\gamma}_f^\dagger \cdot \bm{N}_f^{-1} \cdot \bm{\gamma}_f\,,
\end{aligned}
\end{equation}
where, matrix $\bm{N}_f$ is the covariance matrix of the CSD~\cite{LazzariniOWF,Ain_2015,PhysRevD.104.022004}. The dirty map represents the sky map as observed through the detectors. By construction, the Fisher information matrix, $\bm{ \Gamma}_f$, serves as the noise covariance matrix for the dirty map (in the weak signal limit) and encodes information about the pixel-to-pixel correlation. In terrestrial GWB experiments, inverting $\bm{ \Gamma}_f$ is nontrivial due to the blind spots of the baseline~\cite{Radiometer_Mitra_2008,Panda19,Agarwal_2021,Xiao23}.  

The ORF exhibits time translation symmetry with a sidereal period, which is leveraged to fold the entire observing run data (spanning around one hundred live days) into a single sidereal day~\cite{Thrane_Folding_2015,Ain_2015}. Additionally, \pyStoch{}~\cite{Ain_2018} is a code that fully utilizes the folded data, enabling the calculation of maps on a personal computer. This approach significantly speeds up the processing, making it hundreds of times faster, and eliminates the need to store large intermediate data files.

The folding and \pyStoch{} code were used to create narrowband maps using data up to the O4a observing run~\cite{LIGOScientific:2025bkz}. The ASAF radiometer search was then performed using these maps to identify persistent sources of GWs~\cite{ASAF_O1-O3}.

In the next section, we will briefly discuss the relationship between narrow band GWB and CW signals.

\subsection{CW source and GW radiometer}

Both CW signals and GWB signals are expected to be persistent over time, but they differ in the length of time coherence. The cross-correlation-based search performed here is suboptimal for CW signals, as it does not fully exploit the long-term coherence, unlike other cross-correlation-based searches discussed in the works of~\citet{Sanjeev2008} and~\citet{Whelan2014}. In contrast, the optimal searches require modeling the phase evolution of the signal over time. Predicting the phase evolution for unknown CW sources, or for sources with frequent frequency drift (such as glitching neutron stars or accreting neutron stars in binaries), demands covering a vast parameter space. In such cases, the ASAF radiometer search, while suboptimal, is a robust alternative due to its insensitivity to the signal model.

A comparison of the sensitivity of the radiometer search with matched-filtering-based searches was presented in~\citet{Messenger2015}, where a mock data challenge for Scorpius X-1 was performed. The study estimated that the radiometer search has a similar sensitivity to TwoSpect~\cite{Goetz_2011} and less sensitivity than CrossCorr search~\cite{Sanjeev2008}. At the same time, the radiometer search uses less than 1\% resources than the other two searches. 
Additionally, a CW signal may have a polarized state depending on the Earth-source geometry (i.e., the inclination and polarization angles). The radiometer search estimates, assuming an unpolarized signal, could be biased for a polarized CW signal~\cite{Sanjeev2008,LIGOScientific:2025bkz}.

\subsection{Full-sky map detection statistic implemented in ASAF search}

We now discuss the previous detection statistic employed to distinguish GW signals from detector noise and identify candidates for follow-up~\cite{ASAF_O1-O3,LIGOScientific:2025bkz}, along with their associated drawbacks.

In the ASAF search, sky maps are created by dividing the sky into 3072 pixels for each frequency bin with a width of $1/32$ Hz bin width. Data from the 20-1726 Hz frequency range are analyzed.

To avoid the (ill-conditioned) Fisher information matrix inversion problem, only the diagonal part of the matrix was used to create sky maps for $\bm{\hat{\mathcal{P}}}_f$, i.e.,
\begin{equation}
    \hat{D}_\alpha(f)\equiv\frac{X_\alpha(f)}{\Gamma_{\alpha\alpha}(f)}\,,
\end{equation} 
and its uncertainty and SNR were calculated using
\begin{equation}\label{eq:defSNR}
\begin{aligned}
    \sigma_\alpha(f)&=[\Gamma_{\alpha\alpha}(f)]^{-1/2}\,,\\
\hat{\rho}_\alpha(f) &= \frac{\hat{D}_\alpha(f)}{\sigma_\alpha(f)}\,.
\end{aligned}
\end{equation}
The SNR is expected to follow a Gaussian distribution with mean zero in the presence of only Gaussian noise~\cite{ASAF_O1-O3}, and a nonzero mean in the presence of a GW source. 

Next, the SNR is used as the detection statistic to test the null hypothesis that the data consist of pure Gaussian noise. A candidate frequency-pixel pair with an SNR exceeding the threshold corresponding to a 5\% global $p$ value (accounting for trial factors) is identified as an outlier and selected for further follow-up. If no frequency-pixel pair reaches this significance the null hypothesis is accepted at this stage. If one or more outliers are identified, they are subjected to further follow-up tests to assess potential noise artifacts and to evaluate their consistency with GW source behavior.

In the case that of no outliers are identified, the subthreshold follow-up candidates are identified. These candidates can then be analyzed using more sensitive search methods specifically optimized for CW signals. For example, ~\citet{Knee2024} performed follow-up analyses of subthreshold candidates from the O3 ASAF search~\cite{ASAF_O1-O3}. A similar strategy has been explored
for candidates identified in narrow band radiometer searches for supernova remnants.~\cite{Salvadore:2025lba}. These methods can likewise be applied to subthreshold candidates identified in ASAF analysis.

To identify potential subthreshold candidates while working with a limited computing budget, we consider the distribution of the maximum pixel SNR, $ \rho_{{\rm max}}(f) \equiv \max_\alpha [\hat{\rho}_\alpha(f)] $~\cite{ASAF_O1-O3}. The entire frequency range is divided into 10 Hz bins, over which the sensitivity of the radiometer does not vary significantly. For each 10 Hz bin, a threshold is set on $ \rho_{\rm max}(f) $ corresponding to the 99th percentile of the histogram created using random-time-shifted data (representing the null distribution). Candidates from the zero-lag run with $ \rho_{\rm max}(f) $ above the threshold are then selected for further investigation. Hence, we select approximately three candidates per 10 Hz range, i.e., three candidates per $10^6$ frequency-pixel pairs. 

However, the ``SNR" detection statistic is affected by the oversampling of the sky grid at low frequencies, as discussed in the next section.

\begin{figure}[t]
  \centering
  \begin{tabular}{cc}
    \subcaptionbox{Noise-only Map: 30 Hz}{\includegraphics[scale=0.2]{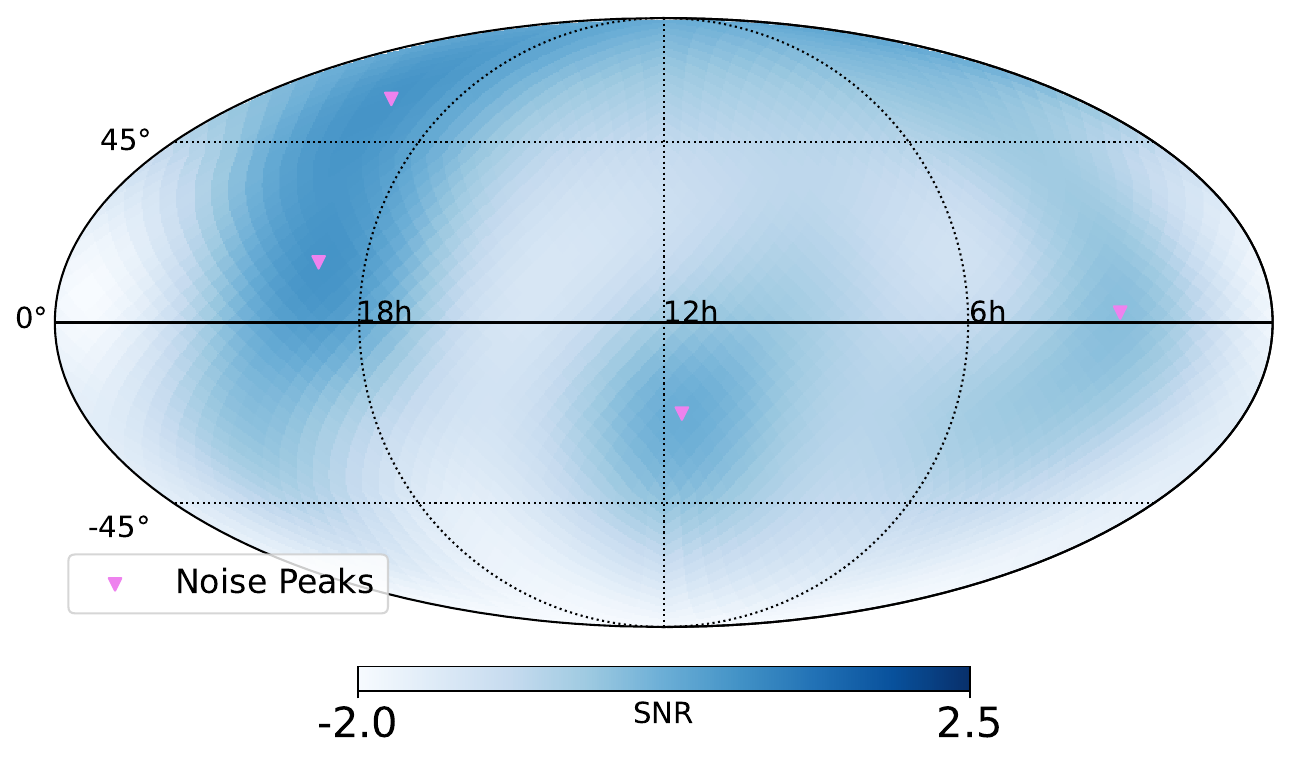}} &
    \subcaptionbox{Noise+injection source: 30 Hz}{\includegraphics[scale=0.2]{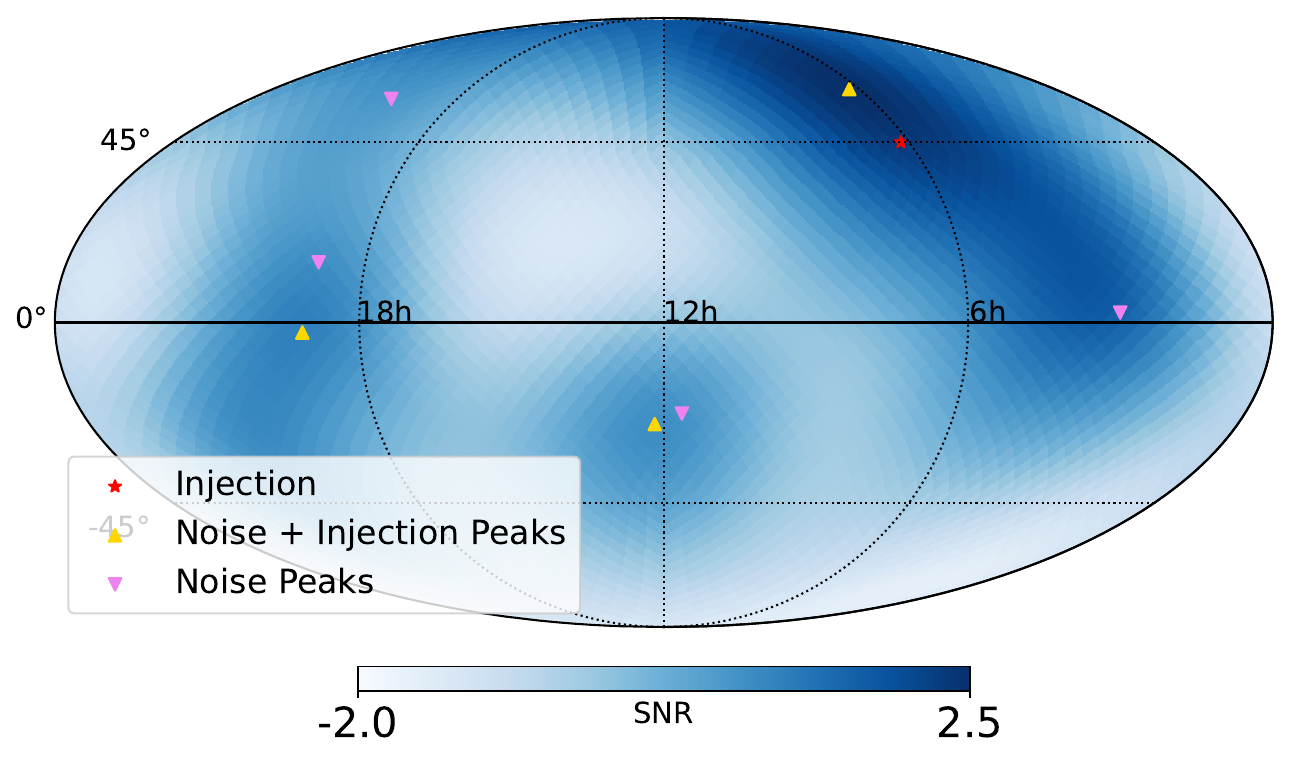}} \\
    \subcaptionbox{Noise-only Map: 200 Hz}{\includegraphics[scale=0.2]{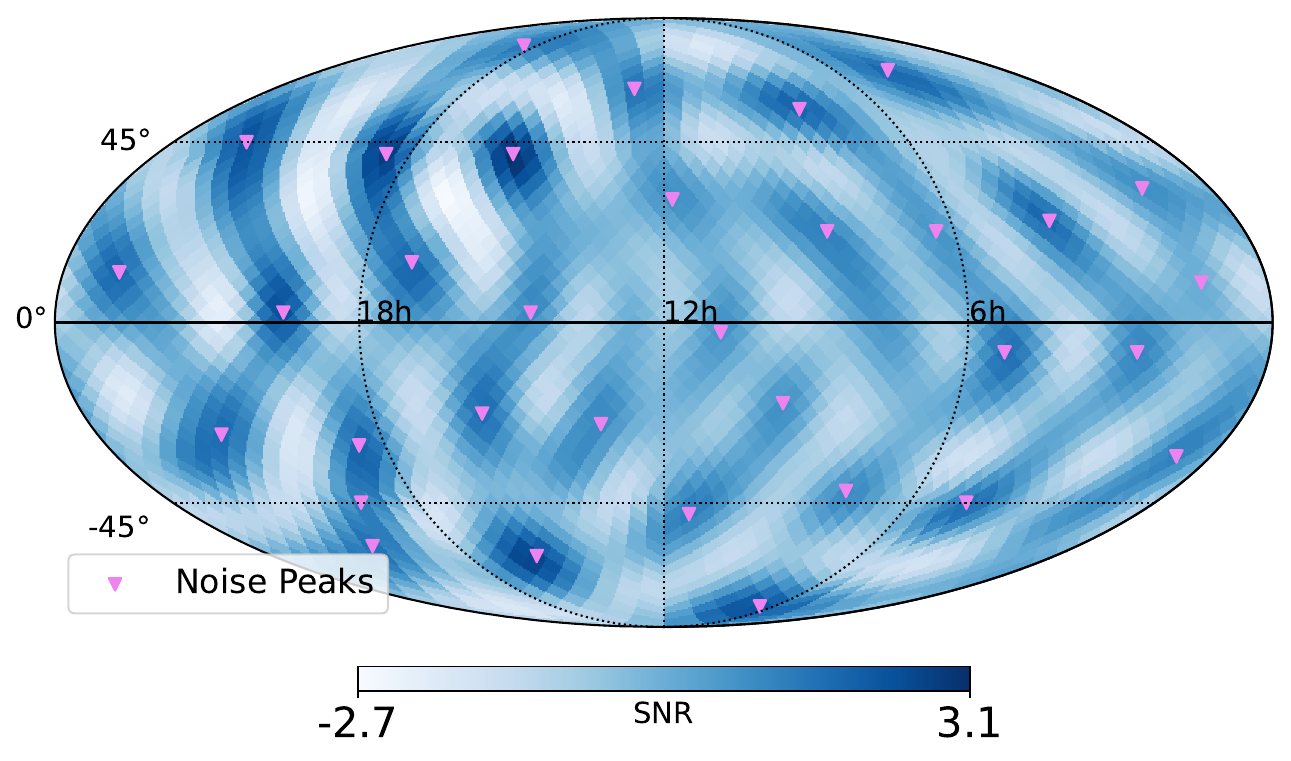}} &
    \subcaptionbox{Noise+injection source: 200 Hz}{\includegraphics[scale=0.2]{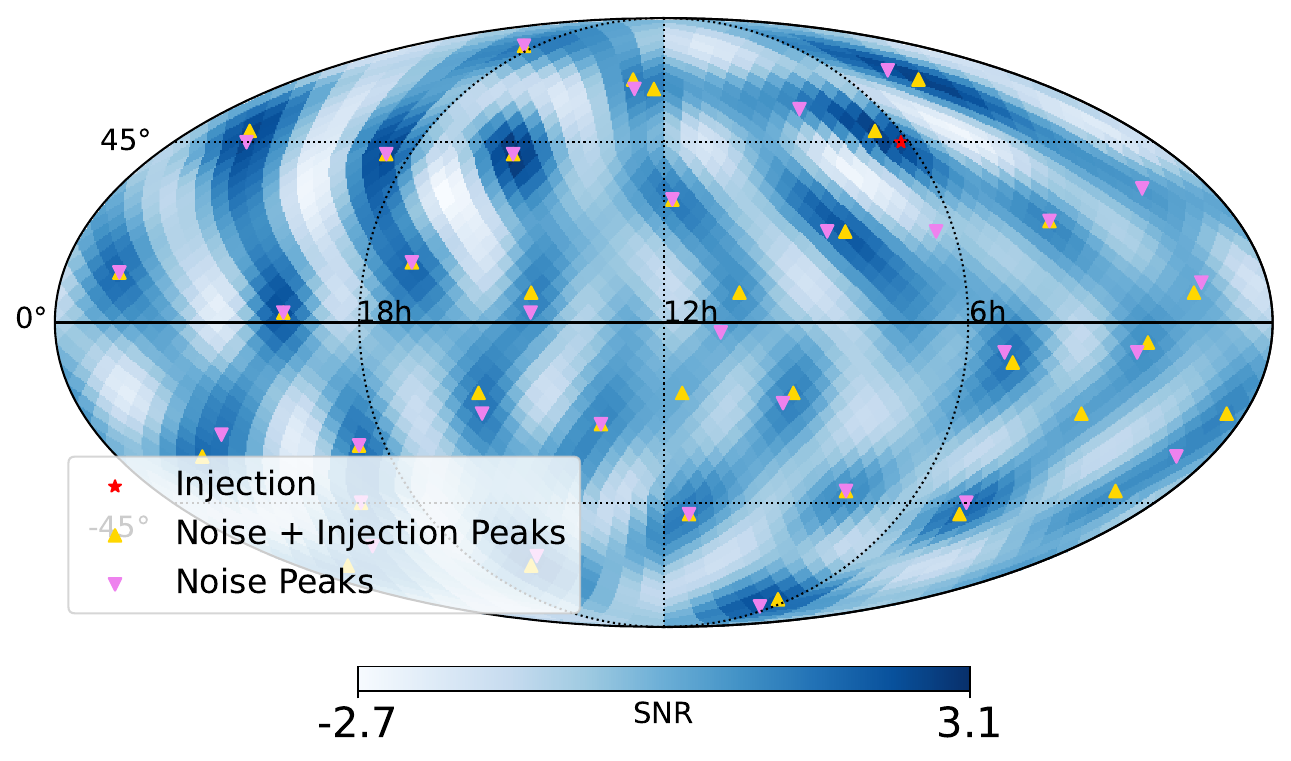}}
  \end{tabular}
  \caption{Simulated SNR sky maps at two frequencies, 30 Hz and 200 Hz, with and without a source. The maps illustrate the point spread function of pixels, showing the spatial distribution of SNR values and the effect of source injection on the observed signal. The source is injected with an SNR of 3 at the sky coordinates (RA, Dec) = $(6.2\,{\rm h}, 45^\circ)$, marked by the red star. The violet downside and yellow upside triangles represent the peaks identified in the noise-only and noise+injection cases, respectively, using the peak finder algorithm introduced in Sec.~\ref{sec:peak_finder}.}
  \label{fig:angularSpread}
\end{figure}

\subsection{Pixel-to-pixel correlation of detection statistic}

The SNR for each frequency-pixel pair, as defined in Eq.~\eqref{eq:defSNR}, serves as the detection statistic. The SNR estimators for different directions at a given frequency can be correlated. The angular spread or leakage depends on the sky direction, signal frequency, and baseline length.

Let us consider two conditions:  
(i) The effect on SNR across the sky given a pointlike GW source and (ii) the SNR across the sky in the case of pure Gaussian noise.

In the first case, the noise-averaged (or noise-GWB-averaged) observed SNR across the sky is proportional to the convolution of the Fisher information matrix and the true SNR sky map (see Appendix~\ref{sec:appendixA}). In the second case, the average SNR is zero, but the noise covariance of the SNR vector is proportional to the Fisher information matrix. Therefore, the Fisher information matrix encodes the correlation information among the pixels across the sky. The second-order statistical properties of the SNR sky map are governed by the Fisher
matrix irrespective of whether a signal is present or not. A naive calculation performed in Appendix~\ref{sec:appendixA} shows that the elements of this matrix are proportional to the ``sinc" function with arguments such as angular distance, frequency, and baseline length. 

The angular spread of the SNR for two example frequencies, with and without a GW source, is shown in Fig.~\ref{fig:angularSpread}. The maps are simulated using methodology described in Sec.~\ref{sec:mapmaking}. As noted, the pixel size used for the search is smaller than the angular distance of correlation; therefore, there are correlated pixels (see Table~\ref{table:size_diff_lim_region} for the number of correlated pixels as a function of frequency). The pixel size was chosen adequate to the most sensitive frequency band ($\sim$ 200 Hz) of the baselines~\cite{Abbott_2021}. Therefore, if there were more than one potential CW candidate with different strengths in a single frequency bin, the top candidates identified in the process described in the previous section may only represent the stronger source. If the stronger candidate is due to noise, and there was a true astrophysical signal which is weaker than the noise, the latter will be missed in the process.

Moreover, because neighboring pixels are correlated, they do not constitute independent statistical trials, and the effective trial is therefore smaller than the naive product of the number of frequency bins and pixels. Overestimating the trial factors may lead to an overconservative SNR threshold, effectively lowering the false alarm rate (FAR) and increasing the false dismissal rate (FDR). In practice, the effective trial factor can be approximated by independent sky positions for each frequency bin or by estimating it empirically using simulations~\cite{LIGOScientific:2025bkz}.

Although the method used in~\citet{LIGOScientific:2025bkz} to identify subthreshold candidates is currently immune to pixel-to-pixel correlation, in the future, if the computing budget allows for following up on more candidates—such as more than one candidate per frequency bin—then it will be necessary to avoid unnecessary follow-up of each correlated sample.

In this article, we propose a peak SNR statistic to group the correlated samples together and identify their representative.

\section{Peak SNR Statistic} \label{sec:peak_finder}

\begin{figure}[t]
\scalebox{0.5}{
\begin{tikzpicture}[node distance=1.5cm]
\node (start) [startstop] {Start};
\node (in1) [io, align=center, below of=start] {SNR Sky Map \\
Indices of 8 Neighbors for Each Pixel};
\node (pro1) [process, align=center, below of=in1] {pix=0\\Peaks=[ ]};
\node (dec1) [decision, align=center,below of=pro1, yshift=-1.5cm] {SNR[pix]\\$\geq$SNR[Neighbors]};
\node (pro2) [process,align=center, right of=dec1, xshift = 3cm] {Append \\ pix to Peaks};
\node (pro3) [process, below of=dec1,yshift=-1.5cm] {pix=pix+1};
\node (dec2) [decision, align=center,below of=pro3, yshift=-0.5cm] {pix $<$ 3072};
\node (ou1) [io, align=center, below of=dec2, yshift=-0.5cm] {Peaks};
\node (stop) [startstop,below of=ou1] {End};

\draw[arrow] (start.south) -- (in1.north);
\draw [arrow] (in1.south) -- (pro1.north);
\draw [arrow] (pro1.south) -- (dec1.north);
\draw [arrow] (dec1) -- node[above] {Yes} (pro2);
\draw [arrow] (dec1) -- node [xshift=10pt]{No} (pro3);
\draw [arrow] (pro2) |- (pro3);
\draw [arrow] (pro3) -- (dec2);
\draw [arrow] (dec2.west) -| node [above,yshift=3cm,xshift=10pt]{Yes} (dec1.west);
\draw [arrow] (dec2) -- node [xshift=10pt]{No} (ou1);
\draw [arrow] (ou1) -- (stop);
\end{tikzpicture}}
\caption{The flow chart illustrates the peak finder algorithm for a single frequency bin. For more details, refer to the main text in Sec.~\ref{sec:peak_finder}.}
\label{fig:flowChart}
\end{figure}
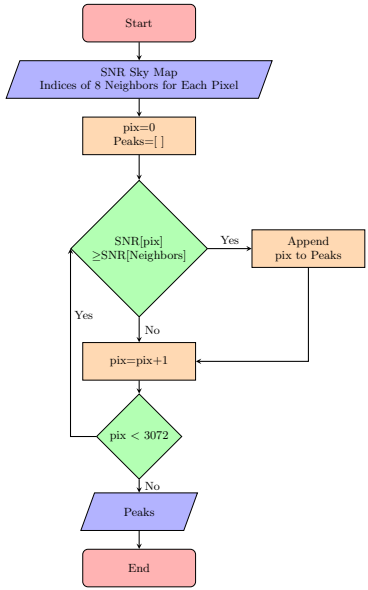

In this section, we describe the working principle of the ``peak finder" algorithm.
The algorithm is presented in the flow chart in Fig.~\ref{fig:flowChart}. We start with the SNR sky map for each frequency and, for each pixel, identify its immediately adjacent neighbors on the \hpx{} grid using the nearest-neighbor relations defined by \hpx{}~\cite{Gorski:2004by}. The procedure begins at the 0th pixel, and we check if its SNR is greater than that of its neighboring pixels. If it is, the pixel is added to the ``peaks" array. Regardless of the outcome, we move to the next pixel and repeat the comparison with its neighboring pixels. This continues for all pixels in the map. At the end, the identified ``peaks" are returned as the output. Examples of the identified peaks in the simulated sky maps are shown in Fig.~\ref{fig:angularSpread}.

\section{Simulation study}\label{sec:sim_study}

We aim to understand the properties of the peak SNR statistic and its effectiveness in enhancing the detection probability of GW sources through Monte Carlo simulations. In this section, we begin by describing the method used to simulate sky maps, followed by an analysis of the results obtained after applying the peak finder algorithm. We present three key sets of results: (i) the reduction in the sample size, (ii) the empirical probability distribution of the statistic and its receiver operating characteristic (ROC) curve, and (iii) the impact on the detection probability of injected sources.

\subsection{Simulating SNR sky maps} \label{sec:mapmaking}

As described in Sec.~\ref{sec:Radiometer}, the dirty map $\bm{X}_f$ and its covariance matrix $\bm{\Gamma}_f$ (or Fisher information matrix) capture sufficient description of the data for our purpose, encapsulating the detector noise and GW signal information compressed from the time domain into the frequency-pixel domain. Therefore, we can leverage the noise covariance matrix $\bm{\Gamma}_f$ in the pixel domain to simulate both pure noise and noise with GW source sky maps.

We use noise sensitivity estimates saved as the folded data from the third LVK observing run, which are publicly available on Zenodo \cite{data}. The folded data covers the frequency range of 20–1726 Hz with a bin width of 0.03125 Hz. Using this data, we compute the Fisher information matrix $\bm{\Gamma}_f$ using the {\tt PyStoch} code, employing a sky resolution of {\tt nside} = 16~\cite{ASAF_O1-O3}.

Next, we fix the injection SNR $\mathcal{\rho}_\alpha(f)$ and compute $\bm{\mathcal{P}}_\alpha(f)$ using the following relation:
\begin{equation}
    \mathcal{P}_\alpha(f) \equiv \rho_\alpha (f) \,\times \sigma_\alpha(f)\,,
\end{equation}
and Eq.~\eqref{eq:defSNR}. For example, if we inject a point source in only the $i$th pixel, then $\rho_i(f)\neq0$, while the SNR remains zero for all other pixels. We then convolve the $\mathcal{P}_\alpha(f)$ sky map with the matrix $\bm{\Gamma}_f$ and add a noise map $\bm{n}_f$, such that
\begin{equation}
\begin{aligned}
    \bm{X}_f &= \bm{\Gamma}_f\,\cdot\,\bm{\mathcal{P}}_f + \bm{n}_f\,,\\
    \bm{n}_f &\sim \mathcal{N}[0,\bm{\Gamma}_f]\,,
  \end{aligned}  
\end{equation}
where $\mathcal{N}[0,\bm{\Gamma}_f]$ represents a multivariate Gaussian distribution. Finally, the SNR map $\mathbf{\hat{\rho}}_\alpha(f)$ is obtained using Eq.~\eqref{eq:defSNR}.

\subsection{Results of peak finder} 

\begin{figure}[t]
    \centering
    \includegraphics[scale=0.3]{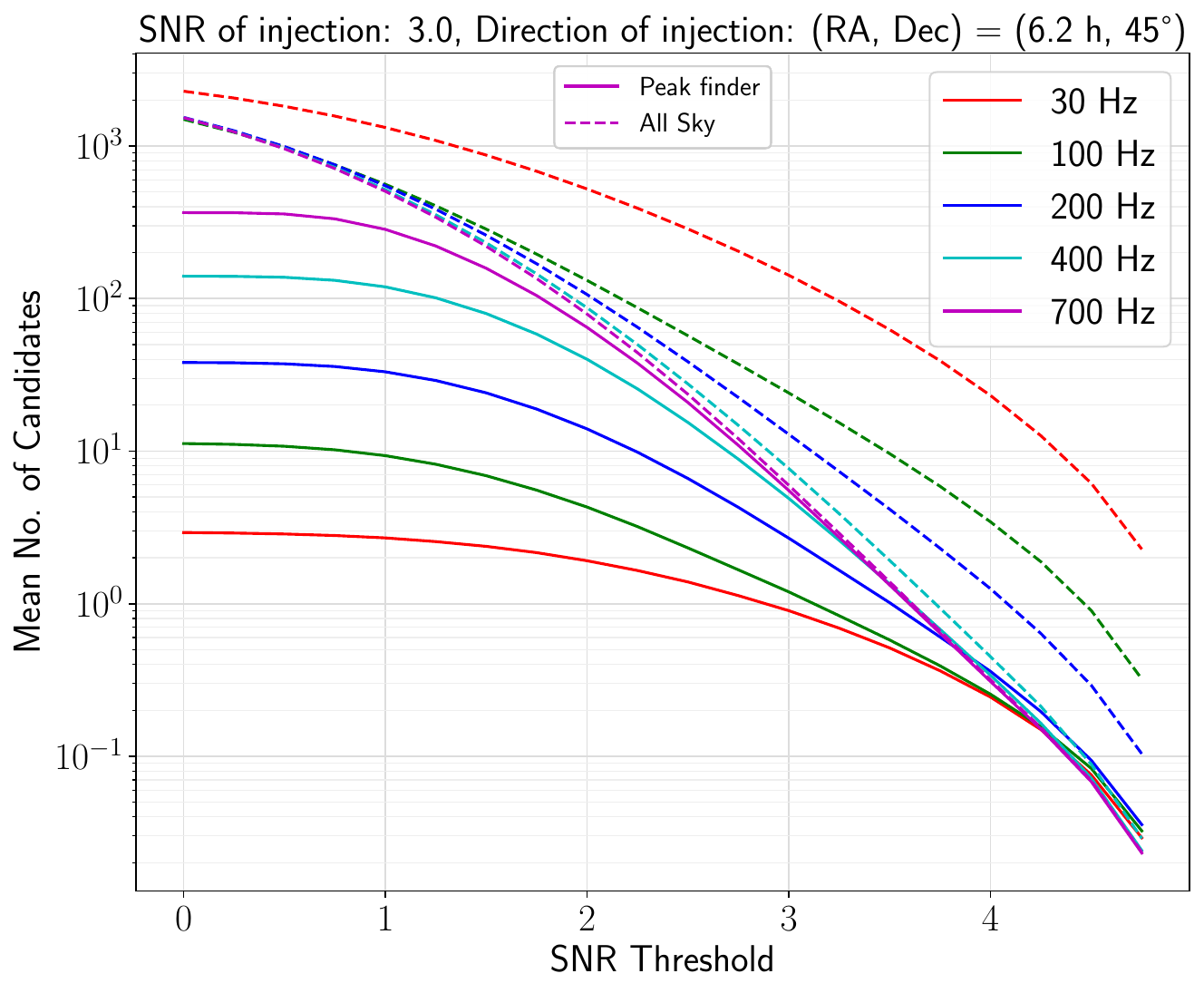}      
    \caption{Mean number of candidates as a function of SNR threshold. Plots are provided for both the all-sky (with dashed line) and peak finder (with solid line) methods across a range of frequencies (30-700 Hz). We note that there is a significant reduction in the number of candidates at lower frequencies with the latter method.}
    \label{fig:NumCandSNR}
\end{figure}

\begin{figure}[t]
    \centering
    \includegraphics[scale=0.45]{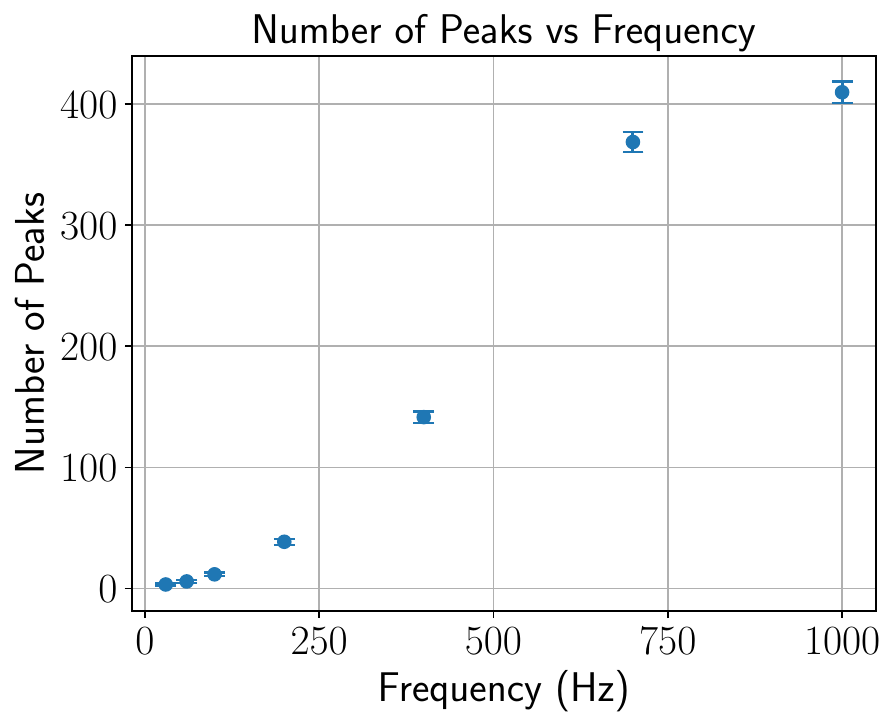}
    \caption{Number of peaks for different frequencies with the mean and standard deviation calculated over $10^4$ maps}
    \label{fig:npeaks_vs_freq}
\end{figure}

\subsubsection{Reduction in sample size}\label{sec:results1}

Simulating SNR sky maps using the method described in the previous section, we apply the peak finder algorithm to identify peaks in the sky map. The identified peaks, for example, at frequencies of 30 and 200 Hz, are shown in Fig.~\ref{fig:angularSpread} for both noise-only and noise+injection source cases. As shown in the figure, for the noise+injection source case, the identified peak is located near the true injection position, though its location exhibits a frequency-dependent uncertainty. This can be clearly seen from Eq.~\eqref{eq:fisher_corr_length}, as the correlation length between pixels is inversely related to frequency. 

Next, we generate $N_{\rm sim} = 10^4$ noise realizations and inject the same GW source to the noise maps. The GW source parameters are fixed with an injection SNR of 3, a fixed sample sky location chosen as (RA, Dec) = $(6.2\,{\rm h}, 45^\circ)$. 
For the ``full-sky" method, a pixel in a realization is considered a candidate if its SNR is greater than the chosen SNR threshold, between 0 and 5. For the peak finder method, we classify a pixel as a candidate if it is a peak and its SNR is greater than the same SNR threshold for each realization. Then the mean number of candidates is defined as $\frac{N_{\rm cand}}{N_{\rm sim}}$, where $N_{\rm cand}$ is the total number of candidates across all realizations. This is repeated for a range of sample frequencies: 30, 100, 200, 400, and 700 Hz.

Figure~\ref{fig:NumCandSNR} shows the reduction in the mean number of candidates with the peak finder algorithm compared to the full-sky map method. The mean number of samples per map above a threshold is significantly reduced at lower frequencies (30–200 Hz). Specifically, at an SNR threshold of 0, the ratio of candidates in the full-sky method to the peak finder method is as follows: (30 Hz: 782:1), (100 Hz: 133:1), (200 Hz: 40:1). At an SNR threshold of 4.75, the corresponding ratios are: (30 Hz: 78:1), (100 Hz: 10:1), (200 Hz: 3:1). However, at higher frequencies (400 and 700 Hz), the reduction factor is smaller, around (10:1 – 1:1) and (4:1 - 1:1), respectively. This is because the point spread function is frequency dependent, and at high frequencies like 400 and 700 Hz, the pixel size is close to or smaller than the diffraction limit. Therefore, there is a much larger number of peaks at such frequencies, as shown in Fig.~\ref{fig:npeaks_vs_freq}. The reduction in sample size is also apparent in sky maps shown in Fig.~\ref{fig:angularSpread}. For example, only three or two peaks are identified in total for maps at 30 Hz. This highlights the significant reduction in the number of samples when applying the peak finder method compared to the full-sky approach.

\begin{figure*}[t]
    \centering
    \includegraphics[width=0.27\textwidth]{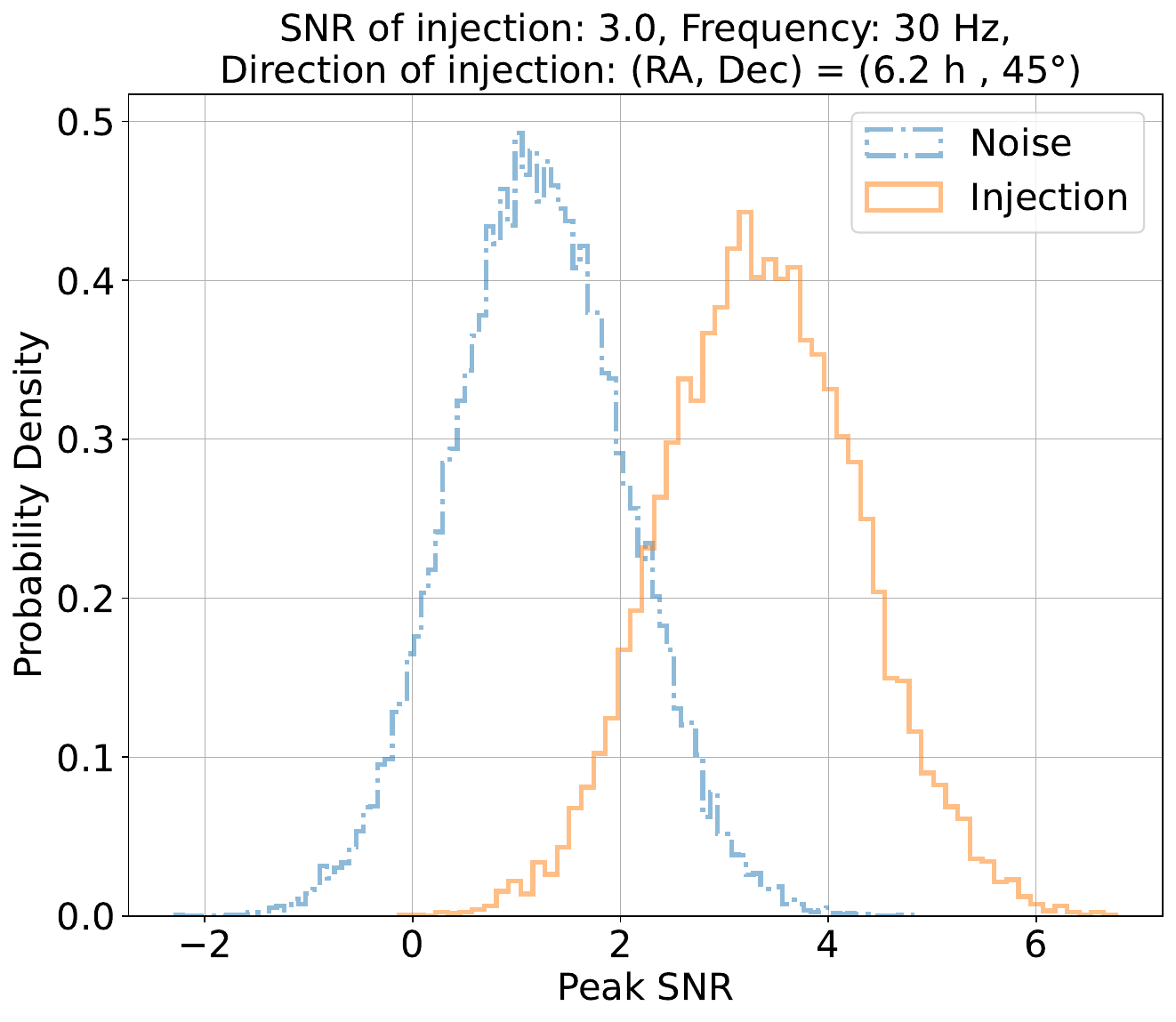}~
    \includegraphics[width=0.3\textwidth]{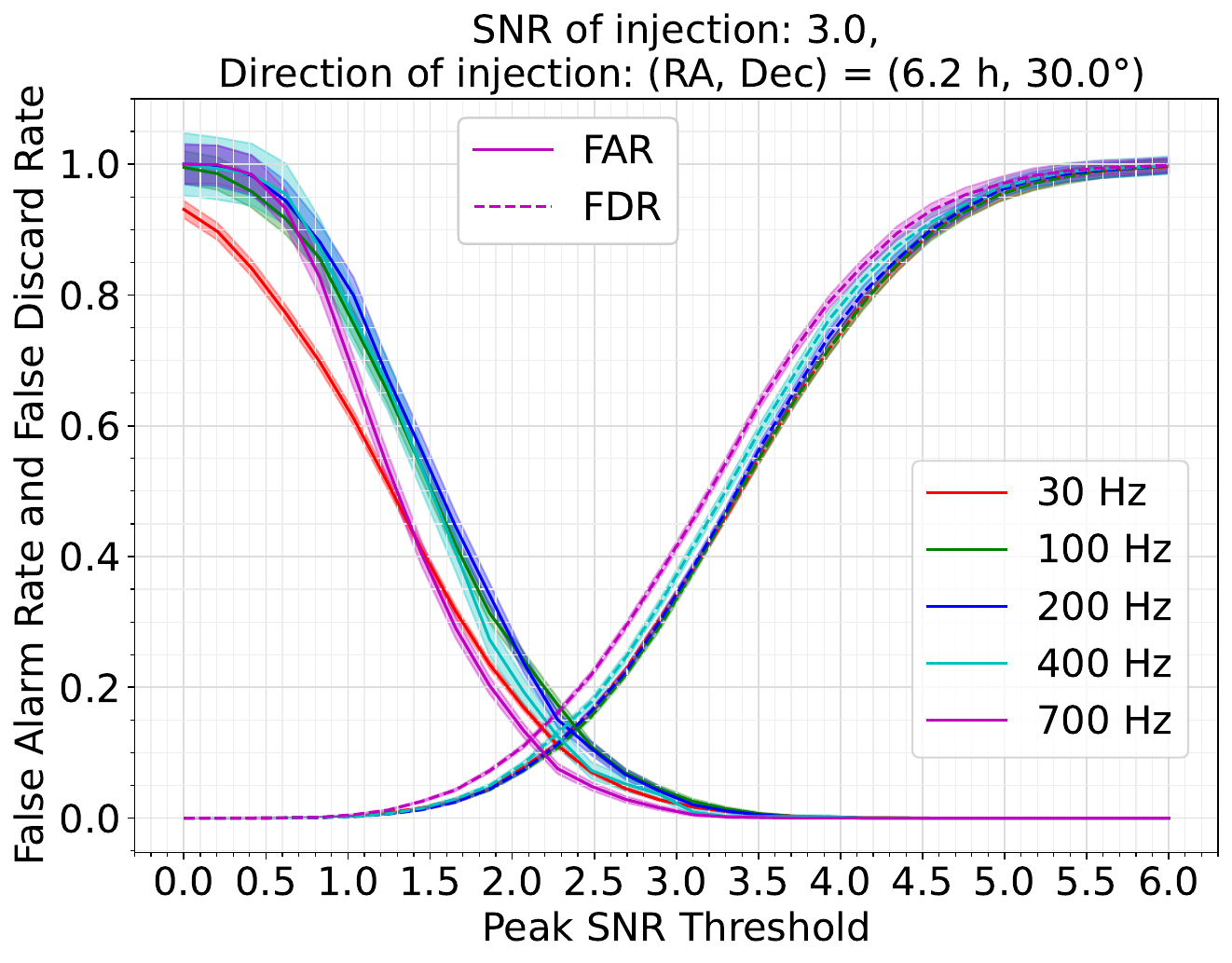}~
    \includegraphics[width=0.3\textwidth]{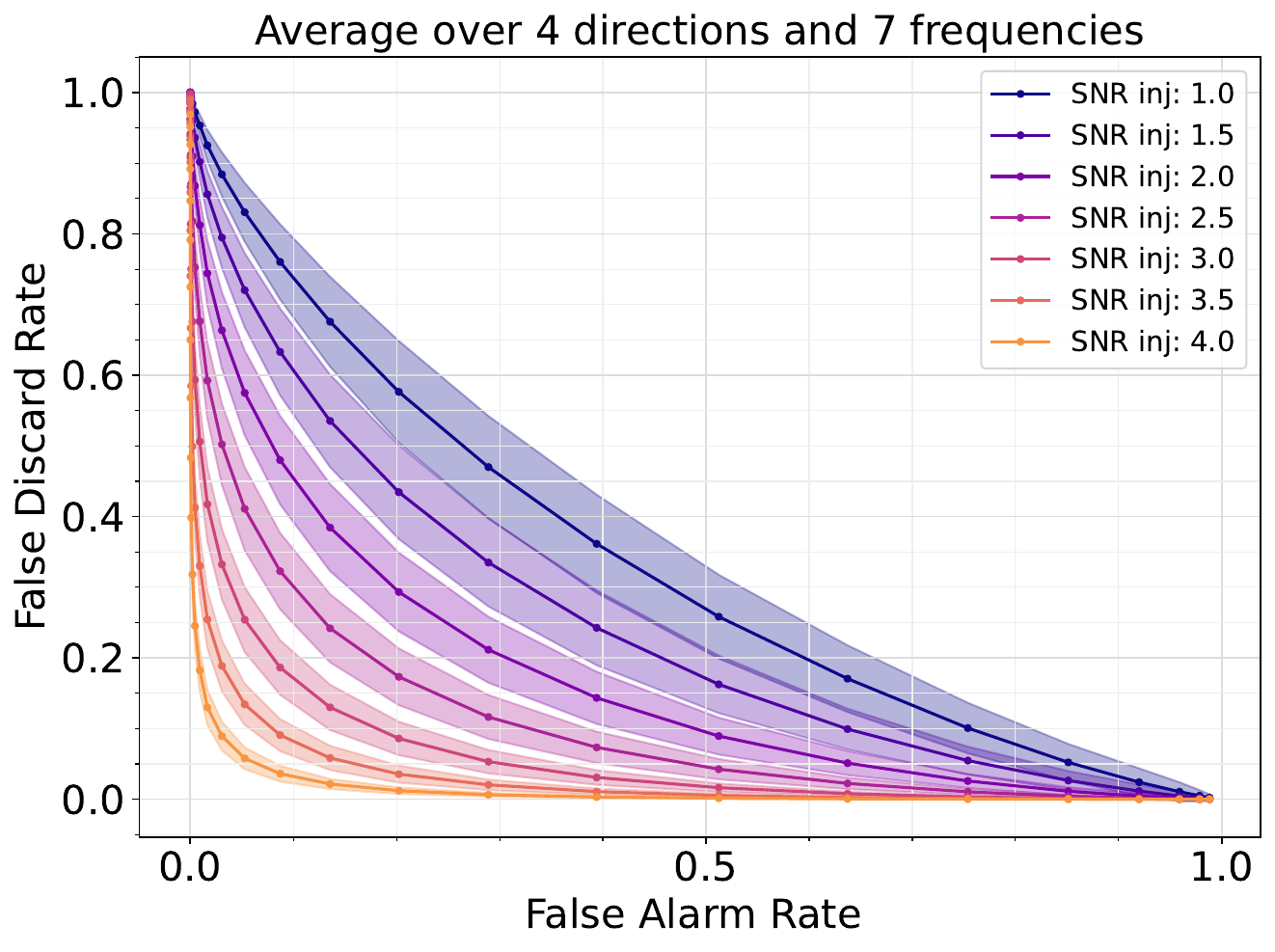}
     \caption{Left panel: The distribution of peak SNR for noise-only (blue dashed) and noise+injection source (orange solid). Middle panel: Variation of FAR (solid) and FDR (dashed) as a function of the threshold on the peak SNR for a set of frequencies. The error bars depict $1\sigma$ Poisson fluctuations in FAR and FDR. Right panel: Frequency-direction-averaged ROC curve for a set of injected SNR values. The threshold decreases from left to right. The error bars represent $1\sigma$ fluctuations across seven frequency bins and four sky directions. See main text in Sec.~\ref{sec:results2} for details.}
    \label{fig:ROC10000_difflim_16AvgDir}
\end{figure*}

\subsubsection{ROC curves}\label{sec:results2}

In the next step, we aim to characterize the probability density function (PDF) of the peak SNR statistic. We use Monte Carlo simulations to empirically estimate the PDF of the peak SNR statistic. Once we have the PDF, we can compute the FAR, and FDR, which are key metrics for evaluating the performance of the peak finder algorithm in identifying GW sources or detection candidates.

Example PDFs for the noise-only and noise+injection source cases are shown in the left panel of Fig.~\ref{fig:ROC10000_difflim_16AvgDir}. The histogram for the noise+injection case is obtained by simulating $10^4$ noise realizations. The GW source parameters are fixed as described in the previous section. We then collect all the peaks that lie within the diffraction-limited region (see Eq.~\eqref{eq:diffraction}) of the injected sky coordinates and plot the histogram of their SNRs. To obtain the noise-only histogram, we repeat the same procedure, but with noise-only maps. This allows us to compare the distribution of the peak SNR statistic in both cases and evaluate how the presence of a GW source affects the statistical behavior of the peaks.

Now, in the next step, we study the effect of the injection SNR on the FAR and FDR. We define our null hypothesis as the data being pure Gaussian noise. The alternative hypothesis is that there is one pointlike GW source present in the data along with Gaussian noise with nonzero power, i.e., $\mathcal{P}_\alpha>0$. The FAR and FDR are then defined as follows:

\begin{equation}
    \begin{aligned}
    \text{FDR} &= 1-\frac{M_{\rm TD}}{M_{ \rm NI}}\,,\\
    \text{FAR} &= \frac{M_{\rm FA}}{M_{\rm  N}}\,,
    \end{aligned}
    \label{eqn:FAR_FDR_def}
\end{equation}
where $M_{\rm TD}$ is the number of noise+injection maps with at least one peak in the diffraction-limited region above the SNR threshold and $M_{\rm NI}$ is number of noise+injection maps with at least one peak in the diffraction-limited region. $M_{\rm FA}$ is the number of noise-only maps with at least one peak in the diffraction-limited region above the SNR threshold and $M_{\rm N}$ is the number of noise-only maps with at least one peak in the diffraction-limited region.

In the middle panel of Fig.~\ref{fig:ROC10000_difflim_16AvgDir}, we show the variation of FDR (dashed) and FAR (solid) with the peak SNR threshold for an example injection with SNR 3 and sky coordinates (RA, Dec) = (6.2 h, $45^\circ$). The error bars represent the Poisson fluctuations around line curves. As expected, FAR decreases with an increasing threshold, while FDR increases. At an SNR threshold of 4, this corresponds to FAR$\sim$0
and FDR$\sim$75\%. FAR and FDR are computed from $10^4$ noise simulations, with the fixed GW source added, based on the definitions in Eq.~\eqref{eqn:FAR_FDR_def}. 

Next, we present the frequency-direction-averaged FAR and FDR for a range of injection SNRs in the right panel of Fig.~\ref{fig:ROC10000_difflim_16AvgDir}. We evaluate this for seven different injection SNRs (ranging from 1 to 4 in steps of 0.5), shown in different colors. The average is performed for four different directions (RA, Dec)= (3 h, 87$^\circ$), (0 h, 2.4$^\circ$), (6.2 h, 30$^\circ$), and (18.2 h, -30$^\circ$) and seven frequencies (30, 60, 100, 200, 400, 700, and 1000 Hz). The error bars represent fluctuations around the solid curve across different frequencies and sky directions. Along each curve, the peak SNR threshold decreases from left to right ranging from 6 to 0. This choice of threshold ensures that FAR equals approximately unity at threshold 0 and the FDR equals unity at threshold 6, providing a useful sanity check. Notably, FDR is dependent on source power and decreases as the injection SNR (or source power) increases. Considering the curve
corresponding to an injection SNR of 3, we find that, after averaging, an SNR
threshold of 4 again gives FAR$\sim$0 and FDR$\sim$75\%.

In hypothesis testing, several criteria exist for determining an ``optimal" threshold on the test statistics to accept or reject hypotheses. The Neyman-Pearson criterion, as outlined in the literature, suggests setting the threshold on the test statistics such that the FDR is minimized while keeping the FAR fixed~\cite{AllenRomano}. {The middle panel of Fig.~\ref{fig:ROC10000_difflim_16AvgDir} can therefore be used to determine an appropriate threshold for the peak SNR statistic. For the example shown in Fig.~\ref{fig:ROC10000_difflim_16AvgDir}, this criteria suggests an optimal SNR threshold is $\sim$2.5, corresponding to a 5\% FAR. 

However in the ASAF search, the alternative hypothesis is composite, allowing for an unknown amplitude (or injection SNR), as well as unknown sky direction, and frequency is unknown. The threshold determined above does not account for these additional uncertainties. A comprehensive study to determine an appropriate SNR threshold for a realistic ASAF search will be explored in future applications to real data.

In the next section, we will demonstrate the improvements that the peak SNR statistic brings to the identification of subthreshold follow-up candidates.

\begin{figure*}[t]
    \centering
    \begin{tabular}{ccc}
        \subcaptionbox{}{\includegraphics[width=0.3\textwidth]{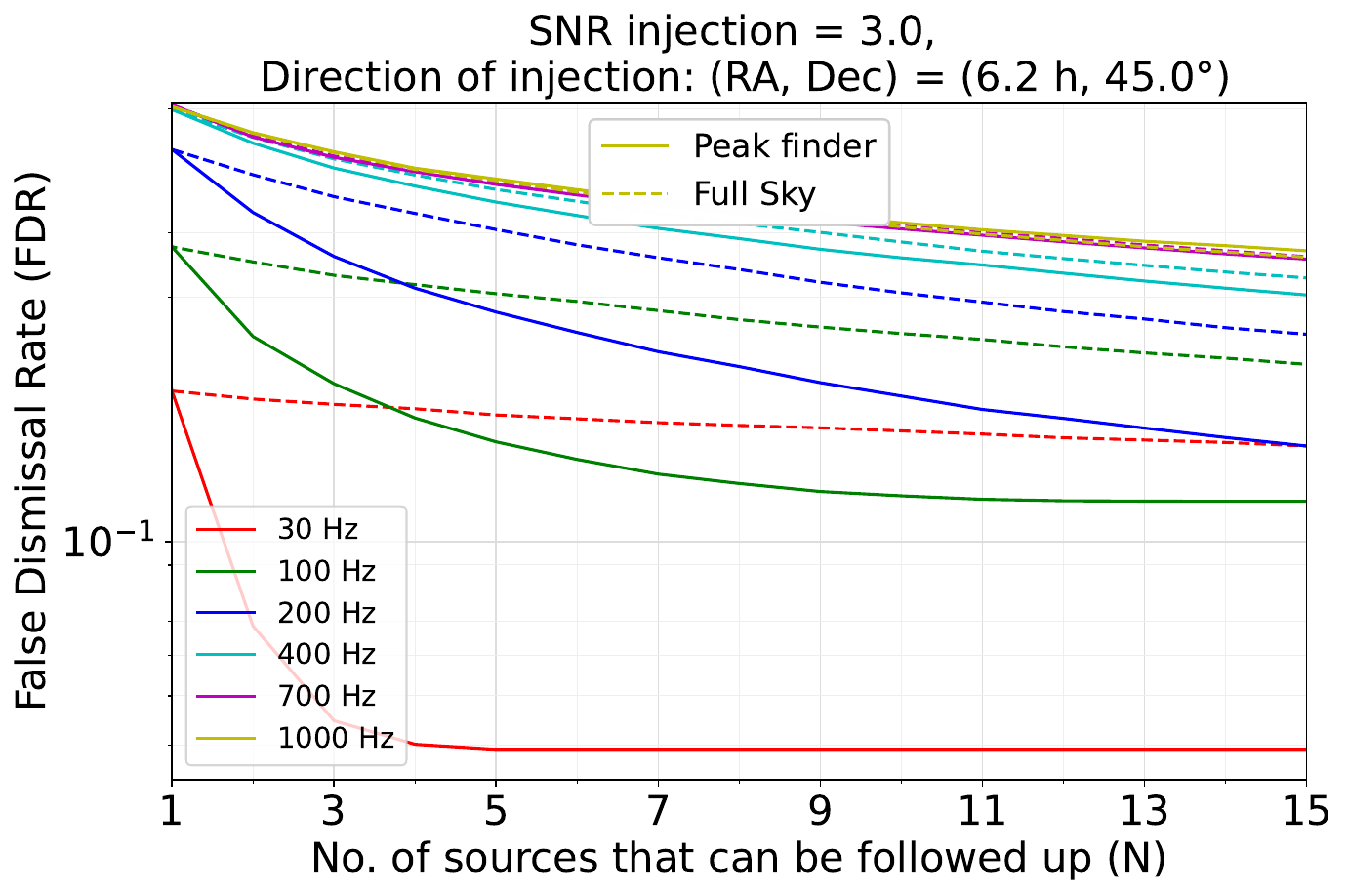}} &
        \subcaptionbox{}{\includegraphics[width=0.3\textwidth]{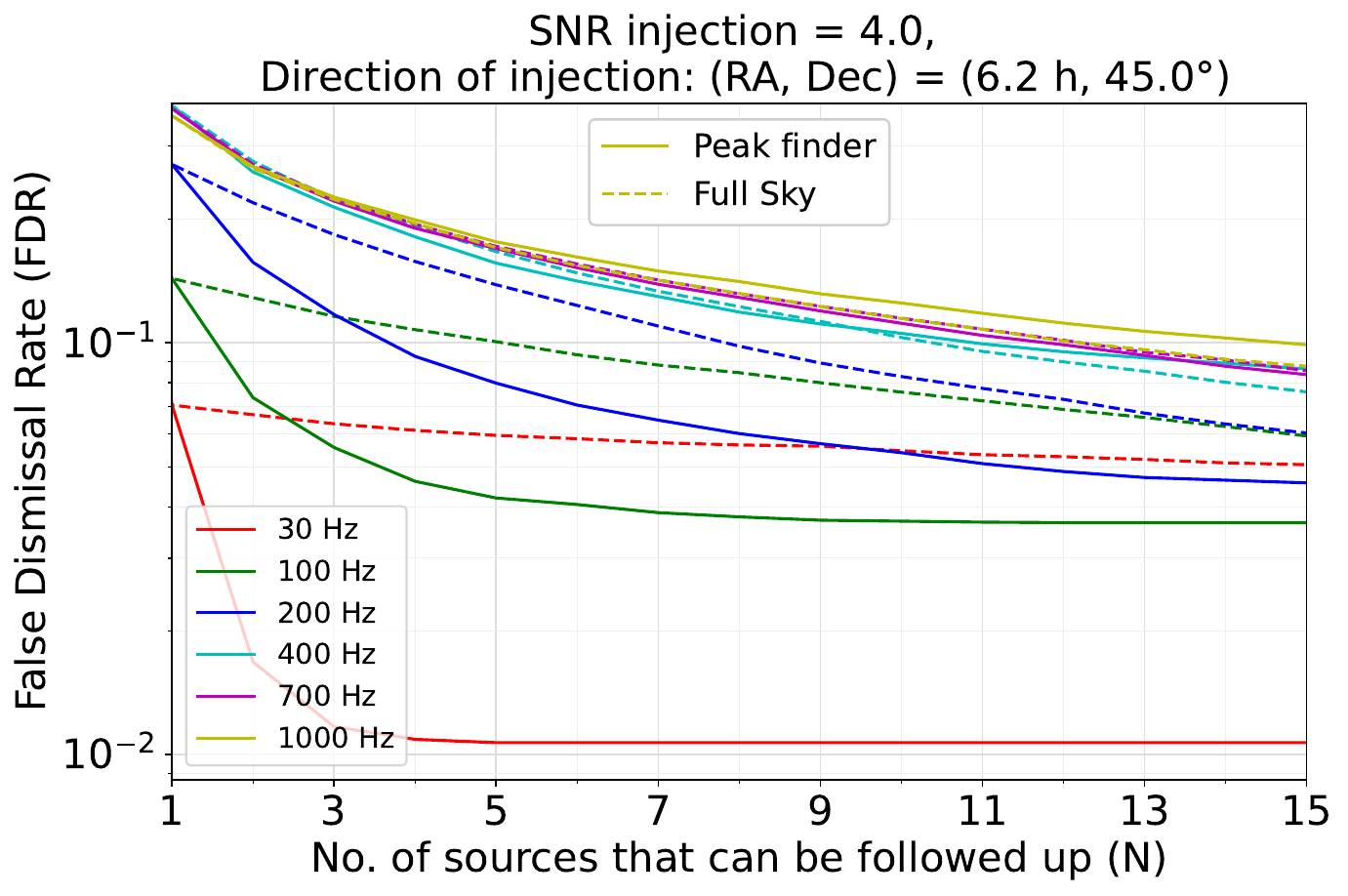}} &
        \subcaptionbox{}{\includegraphics[width=0.3\textwidth]{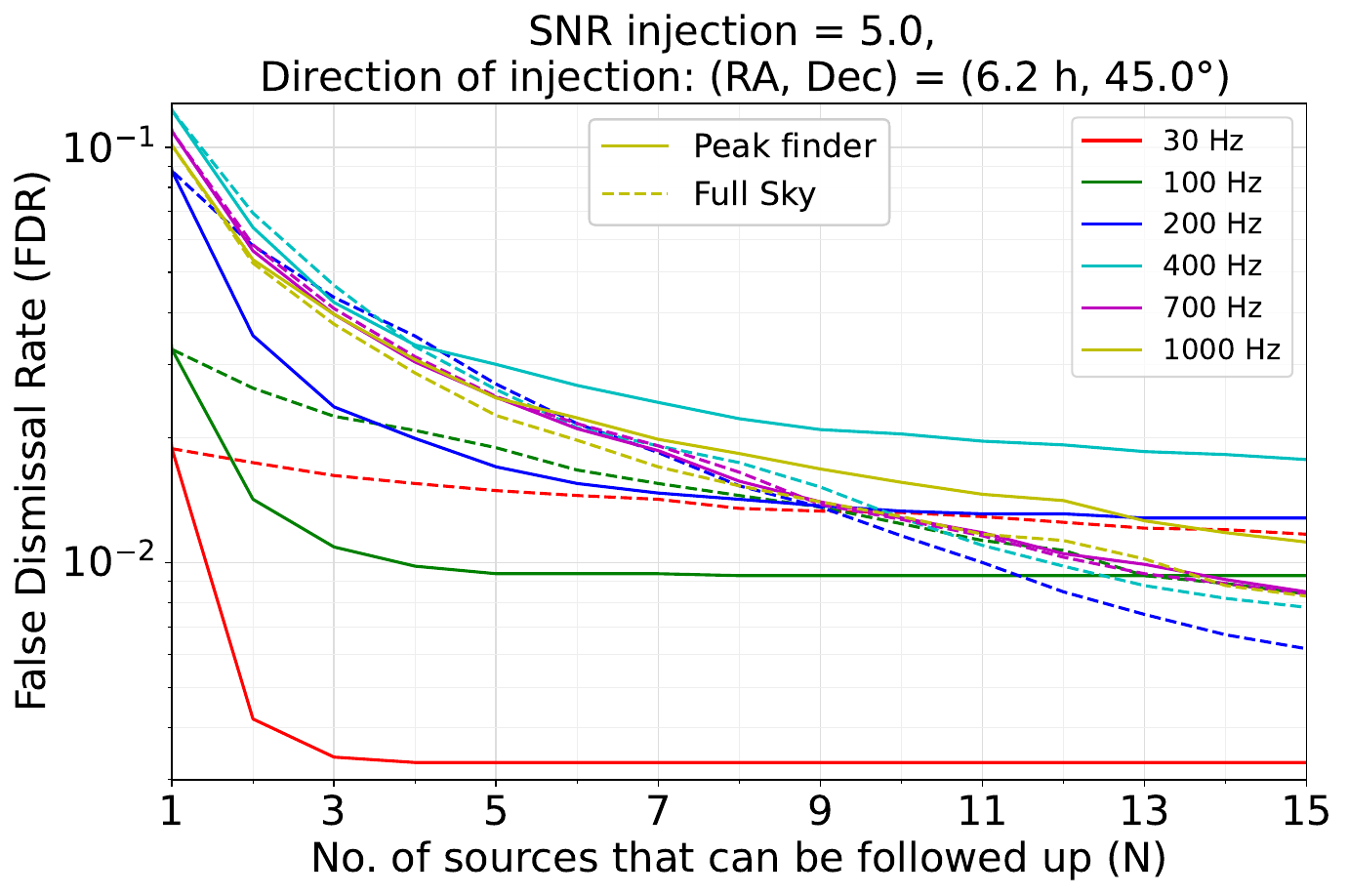}} \\
        \subcaptionbox{}{\includegraphics[width=0.3\textwidth]{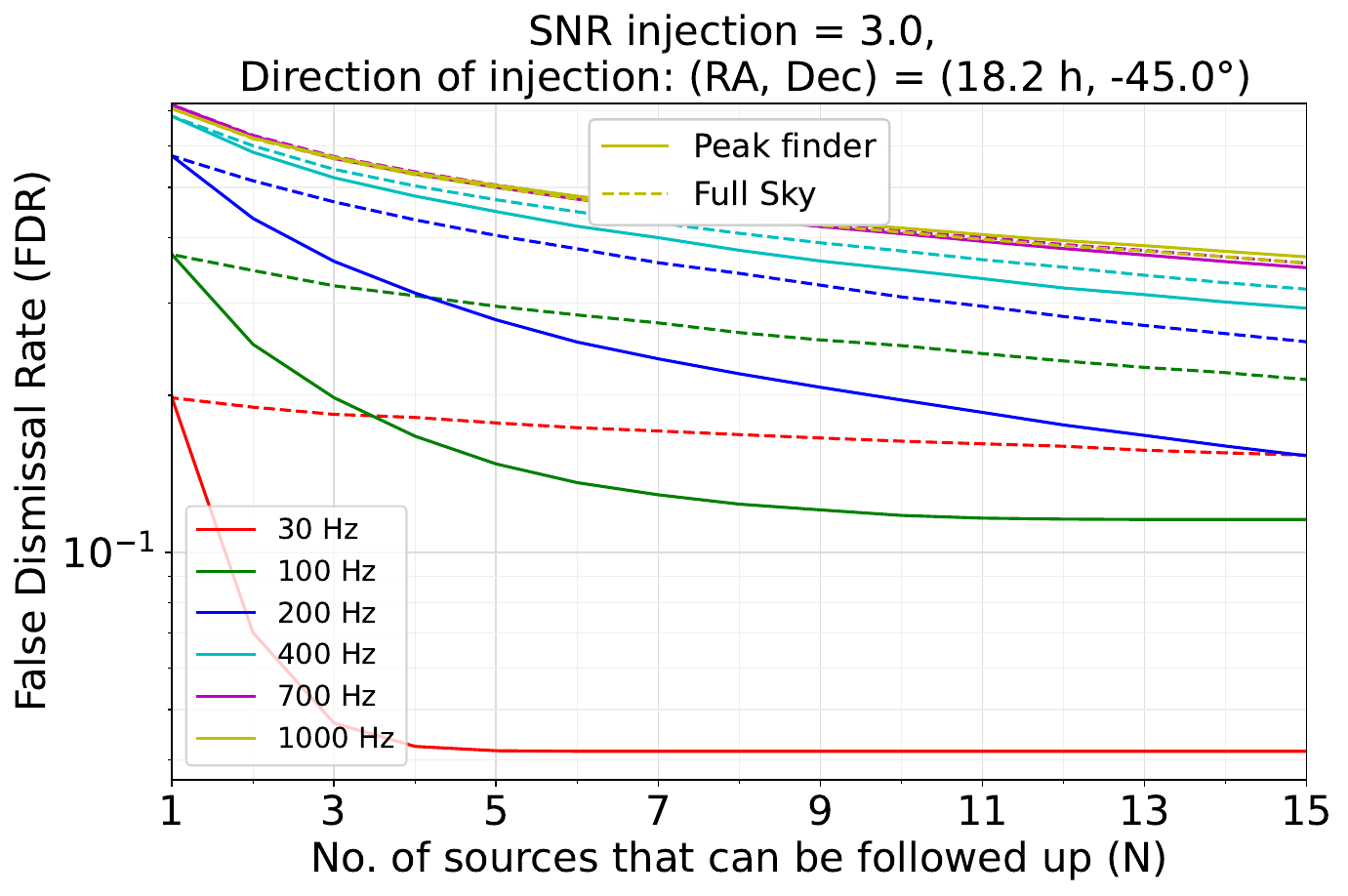}} &
        \subcaptionbox{}{\includegraphics[width=0.3\textwidth]{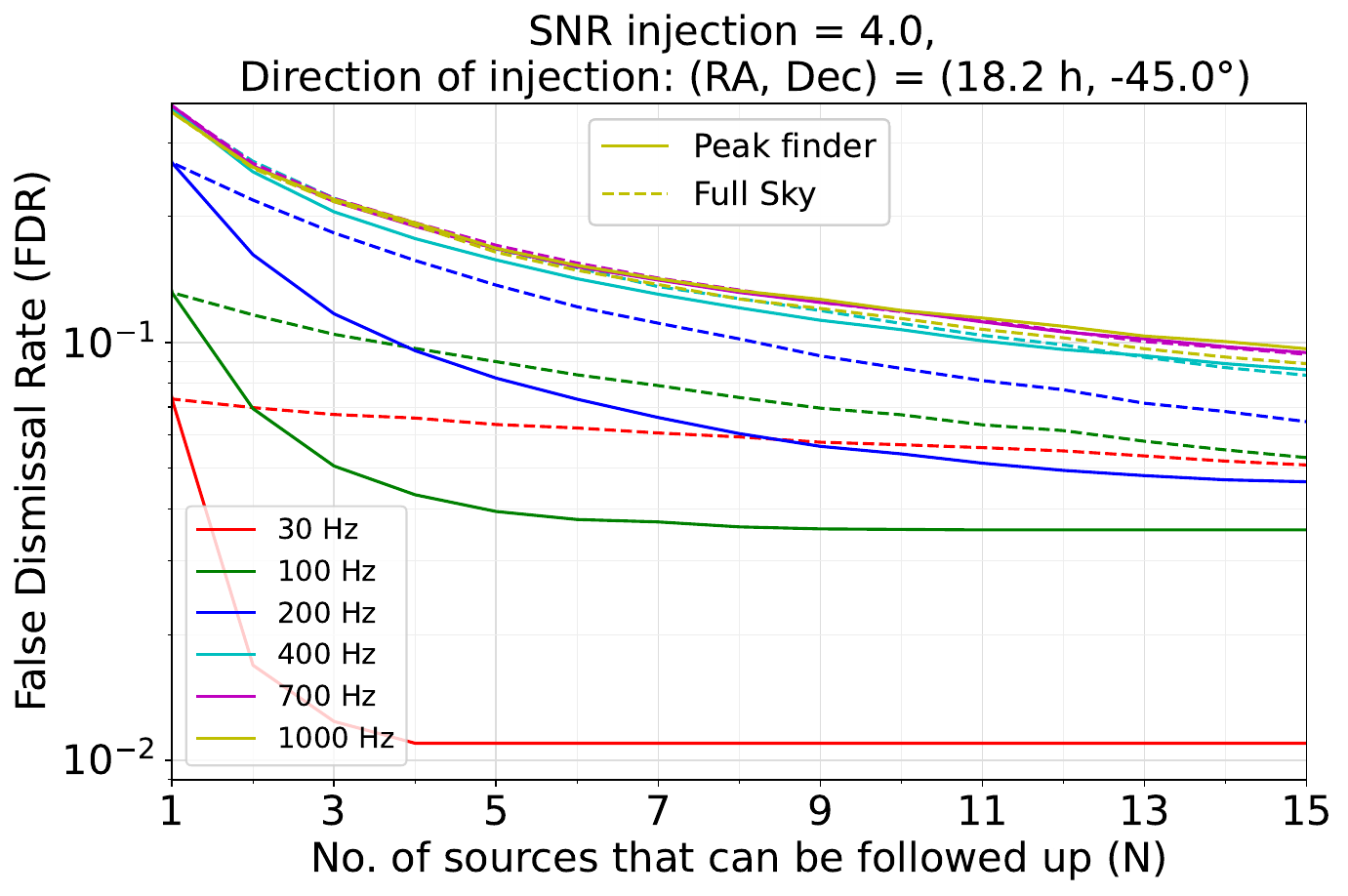}} &
        \subcaptionbox{}{\includegraphics[width=0.3\textwidth]{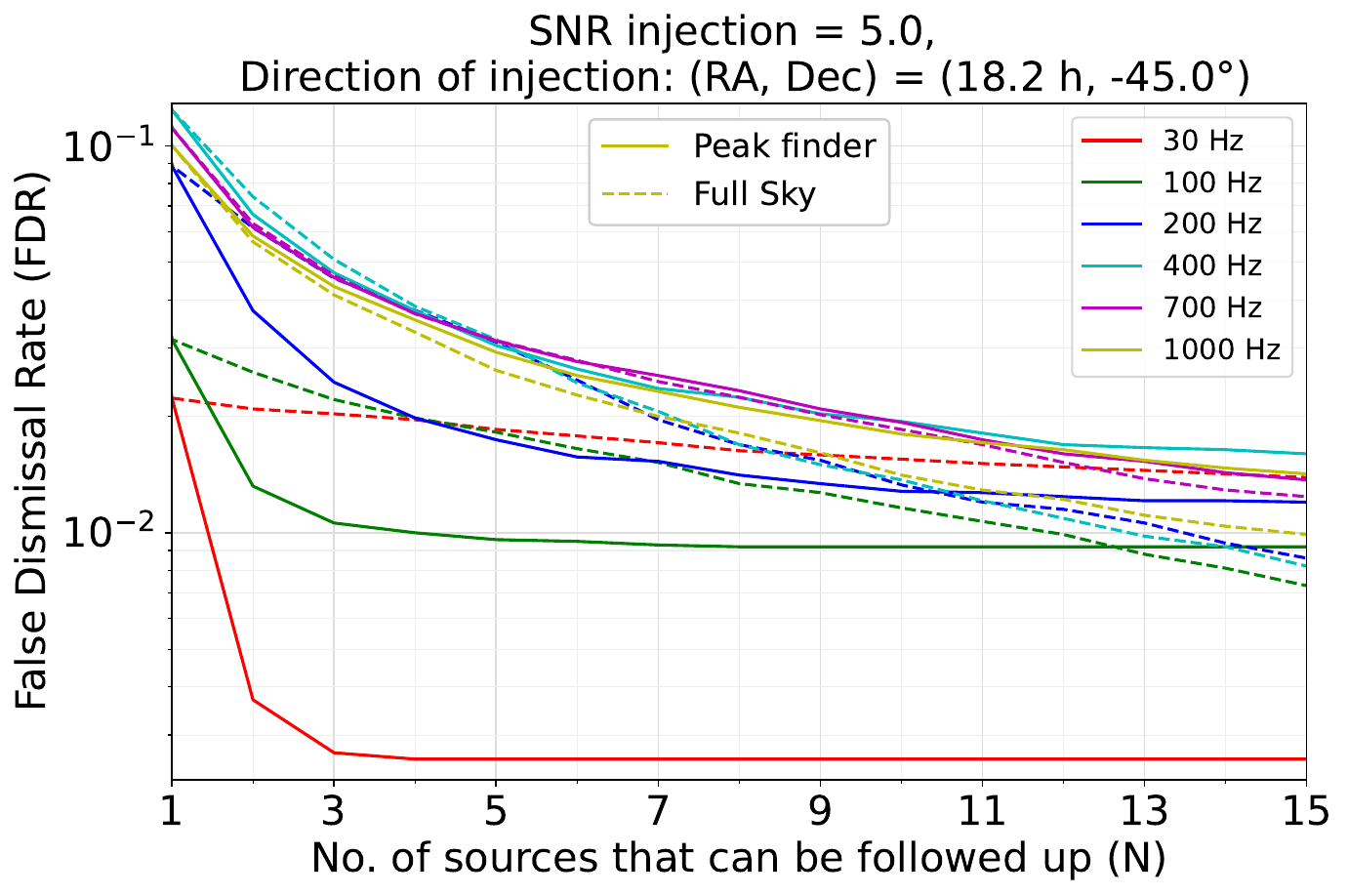}} \\
    \end{tabular}
    \caption{A comparison of the false dismissal rate (FDR) between the full-sky (dashed) and peak-finder (solid) methods as a function of the number of follow-up candidates. In each case, FDR decreases significantly when \( N > 1 \) with the peak-finder method. For example, in the left panel at 30 Hz, FDR is reduced by approximately a factor of 3 for \( N = 2 \). As the frequency increases, the difference in FDR between the two methods diminishes, and beyond a certain point, the trend reverses—FDR for the full-sky method becomes lower than that of the peak-finder method. Moving from the left to the right panel, the injection SNR increases, while the top and bottom rows correspond to injections in two distinct sky directions. As the injection SNR increases, the overall FDR decreases while maintaining the same trend with frequency variations.}
    \label{fig:FDR_numCan}
\end{figure*}

\subsubsection{Impact on FDR with follow-up of subthreshold candidates}

The sensitivity of the ASAF radiometer search can be enhanced by following up on subthreshold candidates using searches optimized for CW signals. In the O3 ASAF search~\cite{ASAF_O1-O3}, 515 candidates were identified~\cite{ASAF_O1-O3}. However, no significant candidates were found after a follow-up using the Viterbi algorithm~\cite{Knee2024}.

In this section, we compare the peak finder method with the full-sky method. The FDR here is defined differently from Eq.~\eqref{eqn:FAR_FDR_def} as ROC curves are not informative if we want to compare the two methods due to two reasons. First, the definition in Eq.~\eqref{eqn:FAR_FDR_def} does not account for the difference in the sample size of the two methods, as shown in Fig.~\ref{fig:NumCandSNR}. For example, at an SNR threshold of 4, the number of candidates is reduced by a factor of $\sim10^2$ at 30 Hz. So, we cannot directly compare them. Secondly, it is very difficult to compare the ROC curves of the two methods since the points of the ROC curves of the two methods with same SNR threshold do not line up together, either along the FAR axis or the FDR axis. Thus, we replace the SNR-based FDR definition with one based on the number of candidates that can be followed up, $N$ Eq.~\eqref{eqn:FDR_def2}. 

In this section, we simulate $N_{\mathrm{sim}}=10^4$ noise realizations and add a fixed GW source with parameters chosen from the combinations mentioned in previous section. In the full-sky method, all pixels in a given realization are treated as candidates, whereas in peak-finder method only peaks in each realization are considered as candidates. For both methods, candidates are first sorted in descending order of their SNR. We then select the top $N$ candidate with the highest SNR in each realization.
If any of those top candidates lies within the diffraction-limited region centered at the injection direction, we consider it a successful detection. Given $N_{\mathrm{det}}$ is the number of realizations with a successful detection, we define the FDR as follows:

\begin{align}
   \text{FDR} & = 1- \frac{N_{\mathrm{det}}}{N_{\mathrm{sim}}}\,.
    \label{eqn:FDR_def2}
\end{align}

In Fig.~\ref{fig:FDR_numCan}, we present the variation of FDR with the number of candidates for a set of frequencies (30, 100, 300, 400, 700, 1000 Hz), injection SNRs (3, 4, 5), and sky directions (RA, Dec) = $(6.2\,{\rm h}, 45^\circ)$ and (18.2 h, -45$^\circ$). The solid curve represents the FDR with peak SNR statistic. The dashed curve corresponds to the FDR computed using the full-sky method. 

We first consider the top-left panel of Fig.~\ref{fig:FDR_numCan} at $ f = 30 $ Hz. For a single candidate ($ N = 1 $), both methods yield identical FDR values. However, as $ N $ increases, the FDR with the peak finder method decreases significantly. For example, at $ N = 2 $, the FDR reduces from 0.19 (full-sky) to 0.07 (peak finder) by nearly a factor of 3. To achieve the same FDR as the peak finder method, the full-sky approach would need to follow up approximately $50$ times more candidates at 30 Hz for an injected signal with SNR$\sim3$, highlighting the substantial reduction in computational cost offered by the peak finder method.

As the frequency increases, the difference in FDR between the two methods starts to decrease up to 200 Hz. At higher frequencies, specifically 400 and 700 Hz, the FDR obtained using the peak SNR statistic becomes higher than that of the full-sky method. The bottom left panel shows similar plots for a different direction (RA, Dec) = (18.2 h, -45$^\circ$). As the injection SNR increases from left to right, the FDR behavior remains the same as for injection SNR = 3, but overall, the FDR decreases as the SNR increases. 

We note that, at each frequency, the FDR obtained with the peak SNR statistic decreases rapidly for small values of $N$ and then saturates. This behavior arises because, once $N$ exceeds the total number of peaks in a given realization, all peaks are necessarily included among the top $N$ candidates. In this regime, a realization is classified as a successful detection provided that at least one peak lies within the diffraction-limited region around the injected signal. Consequently, the FDR reduces to the fraction of realizations that contain no peaks within the diffraction-limited region. This quantity is independent of $N$, leading to the observed flattening of the FDR curves in Fig.~\ref{fig:FDR_numCan}.

In contrast, the FDR for the full-sky statistic decreases more gradually, but continues to decline monotonically. In the limit $N\rightarrow$ (total  number of pixels in a realization), the full-sky FDR approaches zero, since following up all pixels guarantees that at least one pixel lies within the diffraction-limited region around the injection, resulting in a 100\% detection rate. As a result, the peak SNR statistic yields lower FDR than full-sky method at small $N$, while the trend reverses at larger values of $N$. This effect is particularly pronounced for an injection SNR of 5. However, the large-$N$ behavior is not of practical relevance, as realistic searches can only follow up a limited number of candidates per frequency. 

Now we can utilize the peak SNR statistic, which is immune to correlation among sky directions, to identify the top, say $ x\% $ or $ N $, candidates. The reduction in trial factors and the computational savings from identifying candidates linked to the same root cause may allow us to follow up on more uncorrelated samples, ultimately leading to a significant decrease in the FDR. 

If we decide to follow only one candidate per frequency, the number of follow-up candidates identified could be increased using the method described in the~\cite{ASAF_O1-O3,LIGOScientific:2025bkz} search paper. Alternatively, if we allow any number of candidates from any frequency, then $ x\% $ from all frequency-peak pairs, which are immune to correlated samples, can be used to identify candidates following the method prescribed in the previous section.

The improvement in the detection rate in realistic settings, along with its effect on strain sensitivity, will be explored in future work.

\section{Conclusion and Future Prospects}\label{sec:Conclusion}

With the initiation of the first all-sky all-frequency radiometer search, a significant advancement has been made toward a fast and model-independent search for narrow band and persistent GW sources. However, several open questions remain in order to fully utilize the search’s potential. In this work, we addressed the issue of correlated sky samples, which can impact candidate selection and follow-up efficiency.

In this study, we proposed a method to bundle correlated samples together and identify their representative using the peak finder algorithm. We then characterized the statistical properties of the peak SNR statistic through Monte Carlo simulations.

We analyzed the FDR and FAR as a function of the detection threshold on the peak SNR statistic in ASAF radiometer search. We demonstrated that the peak SNR method, which is immune to correlations among sky directions, can effectively reduce trial factors and computational costs while improving candidate selection for follow-up analysis. Our results show that at lower frequencies (e.g., 30 Hz), the peak SNR method significantly reduces the FDR when multiple candidates per frequency bin ($N > 1$) are considered. However, as frequency increases beyond 200 Hz, the difference between the peak SNR and full-sky methods diminishes. This suggests that the optimal follow-up strategy may depend on frequency and signal characteristics. Future work will explore the impact of these findings in realistic search settings, including improvements in detection rates and strain sensitivity.

In this study, we fix the sky-map resolution and assess the performance of the proposed algorithm under this assumption. A systematic analysis of the impact of {\tt nside} on angular resolution and peak localization is beyond the scope of this work. We plan to investigate this dependence in future studies with realistic detector noise and a full end-to-end pipeline.

Here, we have proposed a simple peak finder method to mitigate the issue of correlated sky samples in ASAF radiometer searches. In the future, more sophisticated clustering methods could be developed or applied to further refine candidate selection and improve detection efficiency.

\begin{acknowledgments}

We would like to thank Iuri La Rosa for carefully reviewing this manuscript as part of the LVK internal review process.

This research has made use of folded data obtained from the Zenodo~\cite{data}, provided by the LIGO Scientific Collaboration, the Virgo Collaboration, and KAGRA. This material is based upon work supported by NSF’s LIGO Laboratory which is a major facility fully funded by the National Science Foundation. The authors are grateful for computational resources provided by the Inter-University Center for Astronomy and Astrophysics (IUCAA), Pune, India, and the LIGO Laboratory and supported by National Science Foundation Grants No. PHY-0757058 and No. PHY-0823459. This article has a LIGO Document No. LIGO-P2500691.

A. S. acknowledges the use of the Maple cluster at the University of Mississippi (funded by NSF Grant CHE- 1338056). D. A. acknowledges financial support from the IUCAA, Pune, India, during the initial phase and the Actions de Recherche Concertées (ARC) and Le Fonds spécial pour la recherche (FSR) of the Féderation Wallonie-Bruxelles during the intermediate phase and The University of Texas Rio Grande Valley, Texas, USA, during the final phase of this work. S. M. acknowledges the Department of Science and Technology (DST), Ministry of Science and Technology, India, for the support provided under the esteemed Swarna Jayanti Fellowships scheme. We used numerous software packages such as {\tt NumPy}~\cite{harris2020array}, {\tt SciPy}~\cite{2020SciPy-NMeth}, {\tt healpy} and {\tt HEALPix}~\cite{Górski_2005,Zonca}, {\tt h5py} \cite{h5py}, {\tt PyStoch}~\cite{Ain_2018,PyStoch}, {\tt pandas}~\cite{reback2020pandas,mckinney-proc-scipy-2010}, and {\tt MATPLOTLIB}~\cite{Hunter:2007} in this work.  

\end{acknowledgments}

\section*{DATA AVAILABILITY}

The data that support the findings of this article are openly available~\cite{sharma2026efficient}.

\appendix

\section{Beam Matrix/Point Spread Function and Noise Covariance Matrix}\label{sec:appendixA}
\subsection{Mean and Covariance of SNR}

We suppress the $f$ dependence. We note that SNR is defined as
\begin{equation}
\hat{\rho}_\alpha \equiv \frac{\hat{D}_\alpha}{\sigma_\alpha}\,,
\end{equation}
where $\hat{D}_\alpha=[\Gamma_{\alpha\alpha}]^{-1}\, X_\alpha$, $\sigma_\alpha=[\Gamma_{\alpha\alpha}]^{-1/2}$. 

Let us understand the contribution of GW source and noise to a dirty map, i.e., given~\cite{Radiometer_Mitra_2008}
    \begin{equation}
\begin{aligned}
X_\alpha (f)&= \tau\, \Delta f\ \Re\, \sum_{It}\  \frac{\gamma^{I*}_\alpha (t,f)\,C^I(t,f)}{P_{I_1}(t,f)P_{I_2}(t,f)}\,,\\
   \bm{\Gamma}\equiv \Gamma_{\al\beta} (f)&= \frac{\tau\, \Delta f}{2}\ \Re\, \sum_{It}\  \frac{\gamma^{I*}_\al (t,f)\, \gamma^I_{\beta} (t,f)}{P_{I_1}(t,f)P_{I_2}(t,f)}\,,\\
    \langle X_\alpha (f) \rangle_{h,n} &= \tau\, \Delta f\ \Re\, \sum_{It}\  \frac{\gamma^{I*}_\alpha (t,f)\,\langle C^I(t,f)\rangle_{h,n}}{P_{I_1}(t,f)P_{I_2}(t,f)}\,,\\
    & \propto \tau\, \Delta f\ \Re\, \sum_{It}\  \frac{\gamma^{I*}_\alpha (t,f)}{P_{I_1}(t,f)P_{I_2}
    (t,f)}\,,\\
    &\langle\left[\tilde{h}^*_{I_1} \tilde{h}_{I_2}+\tilde{n}^*_{I_1} \tilde{h}_{I_2}+\tilde{h}^*_{I_1} \tilde{n}_{I_2}+\tilde{n}^*_{I_1} \tilde{n}_{I_2}\right]\rangle_{h,n}\\
    & = \tau\, \Delta f\ \Re\, \sum_{It}\  \frac{\gamma^{I*}_\alpha (t,f)}{P_{I_1}(t,f)P_{I_2}\,
    (t,f)}\,\langle \tilde{h}^*_{I_1} \tilde{h}_{I_2}\rangle_{h,n}\\
    & \equiv \sum_{\beta}\, \Gamma_{\alpha\beta}(f)\, \mathcal{P}_{\beta}(f)\,.
\end{aligned}
\end{equation}
Then, the mean of SNR in the presence of a point source in the direction $\alpha_0$, i.e., $\mathcal{P}_{\alpha}(f) = \delta_{\alpha\alpha_0}\,\mathcal{P}_{\alpha_0}(f)$  is given by,
\begin{equation}
\langle \hat{\rho}_\alpha(f) \rangle_{h,n}= \frac{\Gamma_{\alpha\alpha_0}(f)\, \mathcal{P}_{\alpha_0}(f)}{[\Gamma_{\alpha\alpha}]^{-1/2}}\,.
\end{equation}
Owing to the finite width of $\Gamma_{\alpha\alpha_0}(f)$, the response to a point source is not only localized to pixel $\alpha_0$.

Next, when there is no GW signal in the data then the mean of SNR is zero (in the presence of uncorrelated Gaussian noise). However, the pixel-to-pixel covariance of SNR (in a weak signal and large chunk limit) is given by~\cite{Radiometer_Mitra_2008}
\begin{equation}
\begin{aligned}
\langle \hat{\rho}_\alpha\,\hat{\rho}_{\beta} \rangle_{h,n}&\propto  \frac{1}{[\Gamma_{\alpha\alpha}]^{-1/2}\,[\Gamma_{\beta\beta}]^{-1/2}}\\
&\sum_{II'tt'}\,\frac{\gamma^{I*}_\alpha (t,f)\gamma^{I}_{\beta} (t',f)}{P_{I_1}(t,f)P_{I_2}
    (t,f)\,P_{I'_1}(t',f)P_{I'_2}
    (t',f)}\,\\
    &\langle \tilde{n}^*_{I_1}(t;f) \tilde{n}_{I_2}(t;f)\,\tilde{n}_{I'_1}(t';f) \tilde{n}^*_{I'_2}(t';f)\rangle_{h,n}\\ 
& = \frac{\Gamma_{\alpha\beta}}{[\Gamma_{\alpha\alpha}]^{-1/2}\,[\Gamma_{\beta\beta}]^{-1/2}}\,.
\end{aligned}
\end{equation}
We note that purely due to the map-making method (and finite width of detector beam response), the SNR estimators could have pixel-to-pixel correlation in the presence of only detector noise. 

\subsection{Diffraction limit}

To get a quantitative estimate of pixel-to-pixel correlation, let us assume that detector noise is stationary during a day and detector response is isotropic. Then,
\begin{equation}
\begin{aligned}
\Gamma_{\alpha\alpha'} &\propto \sum_t\,e^{i2\pi f (\hatom_\alpha-\hatom_{\alpha'}) \cdot \Delta\vec{x}_I(t)/c}\,.
\end{aligned}
\end{equation}
Let us make another simplification: the projection of baseline length vector $\Delta\vec{x}_I(t)$ on $(\hatom_\alpha-\hatom_{\alpha'})$ varies from $-L/2$ to $L/2$ as Earth rotates where L is the maximum distance between two detectors. The sum over time segments can be replaced by integration over length; then
\begin{equation}
\begin{aligned}
\Gamma_{\alpha\alpha'} &\propto {\rm sinc}\left[\pi f (|\hatom_\alpha-\hatom_{\alpha'}|) \cdot L/c\right]\,,
\label{eq:fisher_sinc}
\end{aligned}
\end{equation}
and first zero of the $``{\rm sinc}"$ function sets a condition on

\begin{equation}
|\hatom_\alpha-\hatom_{\alpha'}| = \frac{c}{fL}\,.    
\label{eq:fisher_corr_length}
\end{equation}

\section{Algorithm Runtime}

The peak finder algorithm takes between 3 and 5 milliseconds to find the peaks in a map on a laptop,\footnote{Laptop specifications: Apple's ARM-based M1 Pro chip, with ten CPU cores (eight performance cores clocked at 3.2 GHz and two efficiency cores at 2 GHz) and 16 GB of memory.} as shown in Fig.~\ref{fig:runtime_vs_freq}. Here we have plotted the mean runtime and its standard deviation for over $10^4$ maps at each frequency. 
\begin{figure}[h]
    \centering
    \includegraphics[width=\linewidth]{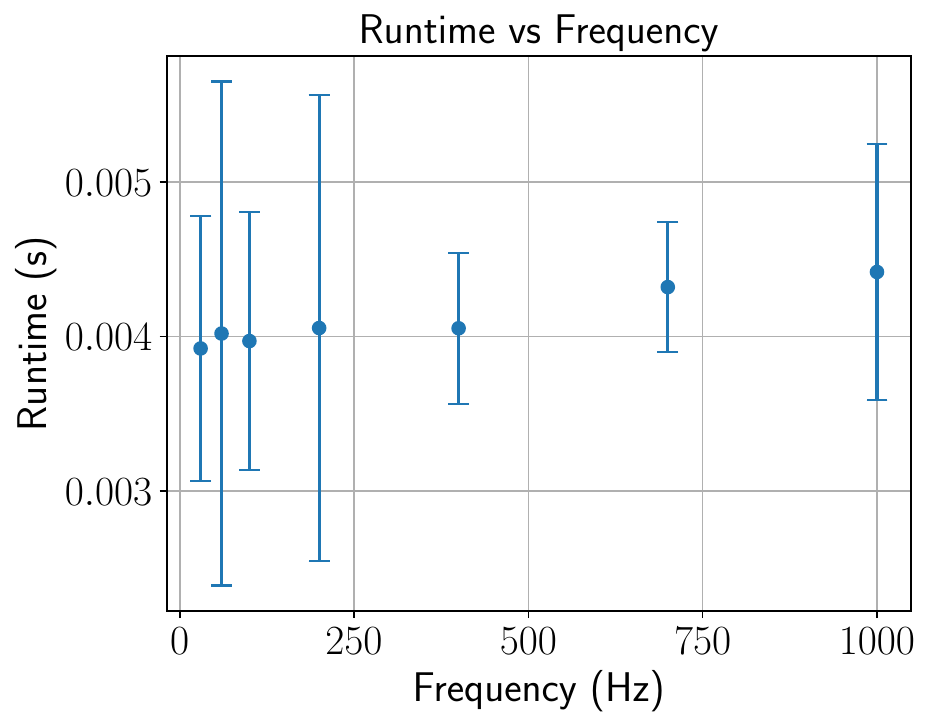}
    \caption{Time taken for the peak finder algorithm to identify peaks as a function of frequency.}
    \label{fig:runtime_vs_freq}
\end{figure}

\bibliographystyle{apsrev4-1}
\bibliography{ref}

\end{document}